\documentclass[nojss]{jss} % article

\usepackage{orcidlink,thumbpdf,lmodern,float}

\usepackage{framed}

\newcommand{\class}[1]{`\code{#1}'}
\newcommand{\fct}[1]{\code{#1()}}

\usepackage[english]{babel}
\author{Sebastian Krantz~\orcidlink{0000-0001-6212-5229}\\Kiel Institute for the World Economy}
\Plainauthor{Sebastian Krantz}

\title{\proglang{collapse}: Advanced and Fast Statistical Computing and Data Transformation in \proglang{R}} % \hphantom{aa}
\Plaintitle{collapse: Advanced and Fast Statistical Computing and Data Transformation in R}
\Shorttitle{\proglang{collapse}: Advanced and Fast Data Transformation in \proglang{R}}

\Abstract{
\pkg{collapse} is a large \proglang{C/C++}-based infrastructure package facilitating complex statistical computing, data transformation, and exploration tasks in \proglang{R}---at outstanding levels of performance and memory efficiency. It also implements a class-agnostic approach to \proglang{R} programming, supporting vector, matrix and data frame-like objects and their popular extensions (\class{units}, \class{integer64}, \class{xts}, \class{tibble}, \class{data.table}, \class{sf}, \class{pdata.frame}), enabling its seamless integration with the bulk of the \proglang{R} ecosystem. This article introduces the package's key components and design principles in a structured way, supported by a rich set of examples. A small benchmark demonstrates its computational performance.
}

\Plainkeywords{statistical computing, vectorization, data transformation and manipulation, class-agnostic programming, summary statistics, C/C++, R}

\Address{
  Sebastian Krantz\\
  Kiel Institute for the World Economy\\
  Haus Welt-Club\\
  D\"usternbrooker Weg 148\\
  24105 Kiel, Germany\\
  E-mail: \email{sebastian.krantz@ifw-kiel.de}
}

\begin{document}

%% -- Introduction -------------------------------------------------------------

%% - In principle "as usual".
%% - But should typically have some discussion of both _software_ and _methods_.
%% - Use \proglang{}, \pkg{}, and \code{} markup throughout the manuscript.
%% - If such markup is in (sub)section titles, a plain text version has to be
%%   added as well.
%% - All software mentioned should be properly \cite-d.
%% - All abbreviations should be introduced.
%% - Unless the expansions of abbreviations are proper names (like "Journal
%%   of Statistical Software" above) they should be in sentence case (like
%%   "generalized linear models" below).

\section[Introduction]{Introduction} \label{sec:intro}
\href{https://sebkrantz.github.io/collapse/}{\pkg{collapse}} is a large \proglang{C/C++}-based \proglang{R} package that provides an integrated suite of statistical and data manipulation functions.\footnote{Website: https://sebkrantz.github.io/collapse/. Linecount (v2.1.2): \proglang{R}: 13,785, \proglang{C}: 19,013, \proglang{C++}: 9,844. \hphantom{aaa.}Exported namespace: 392 objects, of which 238 functions (excl. methods and shorthands), and 2 datasets.}
% Most of these statistical functions are vectorized along multiple dimensions (notably along groups and columns) and perform high-cardinality operations\footnote{With many columns and/or groups relative to data size.} very efficiently.
Core functionality includes a rich set of S3 generic (grouped, weighted) statistical functions for vectors, matrices, and data frames, which provide efficient low-level vectorizations, OpenMP multithreading, and skip missing values by default (\code{na.rm = TRUE}). It also provides powerful data manipulation functions, including vectorized and verbose hash-joins and fast (aggregation, recast) pivots, functions and classes for indexed (time-aware) computations on time series and panel data, recursive tools to deal with nested data, and advanced descriptive statistical tools. This functionality is powered by efficient algorithms for grouping, ordering, deduplication, and matching callable at both \proglang{R} and \proglang{C} levels. The package also includes efficient object converters, functions for memory efficient \proglang{R} programming, such as (grouped) transformation and math by reference, and helper functions to handle variable labels, attributes, and missing data. \pkg{collapse} is \href{https://sebkrantz.github.io/collapse/articles/collapse_object_handling.html}{class-agnostic}, providing statistical operations on vectors, matrices, and data frames/lists, and seamlessly supporting extensions to these objects popular in the \proglang{R} ecosystem such as \class{units}, \class{integer64}, \class{xts}, \class{tibble}, \class{data.table}, \class{sf}, and \class{pdata.frame}. It is globally and interactively configurable, allowing changes to defaults of key function arguments, such as \code{na.rm} arguments to statistical functions or \code{sort} arguments to grouping algorithms (default \code{TRUE}), and modifications of the namespace to mask/replace equivalent but less performant base \proglang{R}/\pkg{tidyverse} functions.\footnote{\pkg{collapse}'s namespace is fully compatible with base \proglang{R} and the \pkg{tidyverse} \citep{rtidyverse}, but can be interactively modified to mask/overwrite key functions with the much faster \pkg{collapse} equivalents. See Section~\ref{sec:glob_opt}.\vspace{-5mm}} % \newline

Why combine all of these features in a package? The short answer is to make computations in \proglang{R} as flexible and powerful as possible. The more elaborate answer is to (1) facilitate complex data transformation, exploration, and computing tasks in \proglang{R}; (2) increase the performance and memory efficiency of \proglang{R} programs; % \footnote{Principally by avoiding \proglang{R}-level repetition such as applying \proglang{R} functions across columns/groups using a split-apply-combine logic, but also by avoiding object conversions and the need for certain classes to do certain things, such as converting matrices to \class{data.frame} or \class{data.table} just to compute statistics by groups.}
and (3) create a new foundation package for statistics and data manipulation that implements many successful ideas in the \proglang{R} ecosystem and other programming environments in a stable, high performance, and broadly compatible manner.\footnote{Such ideas include \pkg{tidyverse} syntax, vectorized aggregations (\pkg{data.table}), data transformation by reference (\pkg{pandas}), vectorized and verbose joins (\pkg{polars}, \proglang{STATA}: \citet{STATA}), indexed time/panel series (\pkg{xts}, \pkg{plm}), summary statistics for panel data (\proglang{STATA}), variable labels (\proglang{STATA}), recast pivots (\pkg{reshape2}), etc...} \newline

\proglang{R} already has a large and tested data manipulation and statistical computing ecosystem. Notably, the \pkg{tidyverse} \citep{rtidyverse} provides a consistent toolkit for data manipulation in R, centered around the \class{tibble} \citep{rtibble} object and tidy data principles \citep{rtidydata}. \pkg{data.table} \citep{rdatatable} provides an enhanced high-performance data frame with parsimonious syntax. \pkg{sf} \citep{rsf} provides a data frame for spatial data and supporting functionality. \pkg{tsibble} \citep{rtsibble} and \pkg{xts} \citep{rxts} provide classes and operations for time series data, the former via an enhanced \class{tibble}, the latter through an efficient matrix-based class. Econometric packages like \pkg{plm} \citep{rplm} and \pkg{fixest} \citep{rfixest} also provide solutions to deal with panel data and irregularity in the time dimension. Packages like \pkg{matrixStats} \citep{rmatrixstats} and \pkg{Rfast} \citep{rfast} offer fast statistical calculations along the rows and columns of matrices as well as faster statistical procedures. \pkg{DescTools} \citep{rdesctools} provides a wide variety of descriptive statistics, including weighted versions. \pkg{survey} \citep{rsurvey} offers computations on complex surveys. \pkg{labelled} \citep{rlabelled} provides tools to deal with labelled data. Packages like \pkg{tidyr} \citep{rtidyr}, \pkg{purrr} \citep{rpurrr} and \pkg{rrapply} \citep{rrapply} provide functions to deal with nested or messy data. \newline

\pkg{collapse} relates to and integrates key elements from these projects. It offers \pkg{tidyverse}-like data manipulation at the speed and stability of \pkg{data.table} for any data frame-like object. It can turn any vector/matrix/data frame into a time-aware indexed series or frame and perform operations such as lagging, differencing, scaling or centering, encompassing and enhancing core manipulation functionality of \pkg{plm}, \pkg{fixest}, and \pkg{xts}. It also performs fast (grouped, weighted) statistical computations along the columns of matrix-like objects, complementing and enhancing \pkg{matrixStats} and \pkg{Rfast}. Its low-level vectorizations and workhorse algorithms are accessible at the \proglang{R} and \proglang{C}-levels, unlike \pkg{data.table}, where most vectorizations and algorithms are internal. It also supports variable labels and intelligently preserves the attributes of all objects, complementing \pkg{labelled}. It provides novel recursive tools to deal with nested data, enhancing \pkg{tidyr}, \pkg{purrr}, and \pkg{rrapply}. Finally, it provides a small but consistent and powerful set of descriptive statistical tools, yielding sufficient detail for most data exploration purposes, requiring users to invoke packages like \pkg{DescTools} or \pkg{survey} only for specialized statistics. \newline

In summary, \pkg{collapse} is a new foundation package for advanced statistical computing and data transformation in \proglang{R} that integrates seamlessly with the ecosystem and offers outstanding speed and memory efficiency. Thus, many core tasks can be done with \pkg{collapse}, and easily extended by specialized packages, yielding more lightweight, faster, and shorter \proglang{R} programs. \newline
% TODO: Need examples of this

Other programming environments such as \proglang{Python} and \proglang{Julia} by now also offer computationally very powerful libraries for tabular data such as \pkg{DataFrames.jl} \citep{jldataframes}, \pkg{Polars} \citep{pypolars}, and \pkg{Pandas} \citep{mckinney2010pandas, pypandas}, and supporting numerical libraries such as \pkg{Numpy} \citep{pynumpy}, or \pkg{StatsBase.jl} \citep{jlstatsbase}. % \pkg{NaNStatistics.jl} \citep{jlnanstatistics}
In comparison with these, \pkg{collapse} offers a class-agnostic approach bridging the divide between data frames and atomic structures, has more advanced statistical capabilities,\footnote{Such as weighted statistics, including several (weighted) quantile and mode estimators, support for fully time-aware computations on irregular series/panels, scaling and centering, advanced (grouped, weighted, panel-decomposed) descriptive statistics etc., all supporting missing values and vectors/matrices/data frames.} supports recast pivots and recursive operations on lists, variable labels, verbosity for critical operations such as joins, and is extensively globally configurable. In short, it is very useful for complex statistical workflows, rich datasets (e.g., surveys), and for integrating with different parts of the \proglang{R} ecosystem. On the other hand, \pkg{collapse}, for the most part, does not offer a sub-column-level parallel architecture and is thus not extremely competitive with top frameworks, including \pkg{data.table}, on aggregating billion-row datasets with few columns.\footnote{As can be seen in the \href{https://duckdblabs.github.io/db-benchmark/}{DuckDB Benchmarks}: \pkg{collapse} is highly competitive on the 10-100 million observations datasets, but deteriorates in performance at larger data sizes. There may be performance improvements for very "long data" in the future, but, at present, the treatment of columns as fundamental units of computation (in most cases) is a tradeoff for the highly flexible class-agnostic architecture.} Its vectorization capabilities are also limited to the statistical functions it provides and not, like \pkg{DataFrames.jl}, to any \proglang{Julia} function. However, as demonstrated in Section~\ref{sec:integration}, vectorized statistical functions can be combined to calculate more complex statistics in a vectorized way. \newline

The package comes with a built-in structured \href{https://sebkrantz.github.io/collapse/reference/index.html}{documentation} facilitating its use. This includes a central \href{https://sebkrantz.github.io/collapse/reference/collapse-documentation.html}{overview page} linking to all other documentation pages and a set of supplementary topic pages which briefly summarize related functionality. The names of these extra pages are collected in a global macro \code{.COLLAPSE\_TOPICS} and can be called directly with \code{help()}:
\begin{Schunk}
\begin{Sinput}
R> .COLLAPSE_TOPICS
\end{Sinput}
\begin{Soutput}
 [1] "collapse-documentation"     "fast-statistical-functions"
 [3] "fast-grouping-ordering"     "fast-data-manipulation"
 [5] "quick-conversion"           "advanced-aggregation"
 [7] "data-transformations"       "time-series-panel-series"
 [9] "list-processing"            "summary-statistics"
[11] "recode-replace"             "efficient-programming"
[13] "small-helpers"              "collapse-options"
\end{Soutput}
\begin{Sinput}
R> help("collapse-documentation")
\end{Sinput}
\end{Schunk}
While this article cannot fully present \pkg{collapse}, the following sections introduce its key features, starting with (\ref{sec:fast_stat_fun}) the \emph{Fast Statistical Functions} and their (\ref{sec:integration}) integration with data manipulation functions; (\ref{sec:ts_ps}) architecture for time series and panel data; (\ref{sec:join_pivot}) table joins and pivots; (\ref{sec:list_proc}) list processing functions; (\ref{sec:summ_stat}) descriptive tools; and (\ref{sec:glob_opt}) global options. Section~\ref{sec:bench} provides a small benchmark, Section~\ref{sec:conclusion} concludes. For deeper engagement with \pkg{collapse}, consult the \href{https://sebkrantz.github.io/collapse/articles/collapse_documentation.html}{documentation resources}, including \href{https://sebkrantz.github.io/collapse/articles/index.html}{vignettes}, \href{https://raw.githubusercontent.com/SebKrantz/collapse/master/misc/collapse\%20cheat\%20sheet/collapse_cheat_sheet.pdf}{cheatsheet}, \href{https://sebkrantz.github.io/Rblog/}{blog}, and \href{https://www.youtube.com/watch?v=OwWT1-dSEts}{talk}$+$\href{https://raw.githubusercontent.com/SebKrantz/collapse/master/misc/useR2022\%20presentation/collapse_useR2022_final.pdf}{slides}.
\section{Fast statistical functions} \label{sec:fast_stat_fun}
The \href{https://sebkrantz.github.io/collapse/reference/fast-statistical-functions.html}{\emph{Fast Statistical Functions}}, comprising \fct{fsum}, \fct{fprod}, \fct{fmean}, \fct{fmedian}, \fct{fmode}, \fct{fvar}, \fct{fsd}, \fct{fmin}, \fct{fmax}, \fct{fnth}, \fct{ffirst}, \fct{flast}, \fct{fnobs}, and \fct{fndistinct}, are a consistent set of S3-generic statistical functions providing fully vectorized statistical operations in R.\footnote{'Vectorization' in \proglang{R}/\pkg{collapse} means that these operations are implemented using compiled \proglang{C/C++} code.} Specifically, operations are vectorized across columns and groups, and may also involve weights or transformations of the input data. Their basic syntax is
\begin{Code}
FUN(x, g = NULL, w = NULL, TRA = NULL, na.rm = TRUE, use.g.names = TRUE, ...)
\end{Code}
with arguments \code{x} - data (vector, matrix or data frame-like), \code{g} - groups (atomic vector, list of vectors, or \class{GRP} object), \code{w} - sampling weights (only some functions), and \code{TRA} - transformation of \code{x}. The following examples with \fct{fmean} demonstrate their basic usage on the familiar \href{https://www.rdocumentation.org/packages/datasets/versions/3.6.2/topics/iris}{\code{iris}} dataset recording 50 measurements of 4 variables for 3 species of iris flowers.\footnote{\fct{num\_vars} returns the numeric variables in a data frame-like object; \fct{cat\_vars} the categorical ones.} All examples support weights (\code{w}), and \fct{fmean} can also be multithreaded across columns (\code{nthreads}).\footnote{Not all functions are multithreaded, and parallelism is implemented differently for different functions, as detailed in the respective function documentation. The default use of single instruction multiple data (SIMD) parallelism also implies limited gains from multithreading for simple (non-grouped) operations.}  % As laid out in the \href{https://sebkrantz.github.io/collapse/articles/collapse_object_handling.html}{vignette on object handling}, statistical functions have basis S3 methods for vectors (\class{default}), \class{matrix}, and \class{data.frame}, which call corresponding \proglang{C} implementations that intelligently preserve object attributes. Thus, the functions can be applied to a broad set of \class{matrix} or \class{data.frame}-based objects without the need to define explicit methods. % Users can also directly call the basis methods in case S3 dispatch does not yield the intended outcome. For example, \code{fmean.default(EuStockMarkets)} computes the mean of the entire matrix. %, and attributes are preserved as much as possible ()
\begin{Schunk}
\begin{Sinput}
R> fmean(iris$Sepal.Length)
\end{Sinput}
\begin{Soutput}
[1] 5.843
\end{Soutput}
\begin{Sinput}
R> fmean(num_vars(iris))
\end{Sinput}
\begin{Soutput}
Sepal.Length  Sepal.Width Petal.Length  Petal.Width
       5.843        3.057        3.758        1.199
\end{Soutput}
\begin{Sinput}
R> identical(fmean(num_vars(iris)), fmean(as.matrix(num_vars(iris))))
\end{Sinput}
\begin{Soutput}
[1] TRUE
\end{Soutput}
\begin{Sinput}
R> fmean(iris$Sepal.Length, g = iris$Species)
\end{Sinput}
\begin{Soutput}
    setosa versicolor  virginica
     5.006      5.936      6.588
\end{Soutput}
\begin{Sinput}
R> fmean(num_vars(iris), g = iris$Species)
\end{Sinput}
\begin{Soutput}
           Sepal.Length Sepal.Width Petal.Length Petal.Width
setosa            5.006       3.428        1.462       0.246
versicolor        5.936       2.770        4.260       1.326
virginica         6.588       2.974        5.552       2.026
\end{Soutput}
\end{Schunk}
\subsection{Transformations}
The \code{TRA} argument toggles (grouped) replacing and sweeping operations (by reference), generalizing \code{sweep(x, 2, STATS = fmean(x))}.\footnote{The \code{TRA} argument internally calls \fct{TRA}: \code{TRA(x, STATS, FUN = "-", g = NULL, set = FALSE, ...)}. \vspace{-12mm}} %Its syntax is
%\begin{Code}
%TRA(x, STATS, FUN = "-", g = NULL, set = FALSE, ...)
%\end{Code}
%where \code{STATS} is a vector/matrix/data.frame of statistics used to transform \code{x}.
Table~\ref{tab:TRA} lists the 11 possible \code{TRA} operations. % to transform \code{x} with the statistics.
\begin{table}[H]
\resizebox{\textwidth}{!}{
  \begin{tabular}{lll}
  \emph{String} && \emph{Description}  \\
  \code{"replace\_na"/"na"}   && replace missing values in \code{x} by \code{STATS} \\
  \code{"replace\_fill"/"fill"}   && replace data and missing values in \code{x} by \code{STATS} \\
  \code{"replace"} && replace data by \code{STATS} but preserve missing values in \code{x} \\
  \code{"-"}   && subtract \code{STATS} (center) \\
  \code{"-+"}  && subtract \code{STATS} and add overall average statistic \\
  \code{"/"}   && divide by \code{STATS} (scale) \\
  \code{"\%"}     && compute percentages (divide and multiply by 100) \\
  \code{"+"} && add \code{STATS} \\
  \code{"*"} && multiply by \code{STATS} \\
  \code{"\%\%"} && modulus (remainder from division by \code{STATS}) \\
  \code{"-\%\%"} && subtract modulus (make data divisible by \code{STATS})
  \end{tabular}
}
\caption{\label{tab:TRA} Available \code{TRA} argument choices.}
\end{table}
For example, option \code{TRA = "fill"} replaces elements with the corresponding statistics:
\begin{Schunk}
\begin{Sinput}
R> fmean(iris$Sepal.Length, g = iris$Species, TRA = "fill")[c(1:5, 51:55)]
\end{Sinput}
\begin{Soutput}
 [1] 5.006 5.006 5.006 5.006 5.006 5.936 5.936 5.936 5.936 5.936
\end{Soutput}
\end{Schunk}
%\code{TRA()} is called internally in the \emph{Fast Statistical Functions}, the \code{TRA} argument is passed to \code{FUN}. Thus \code{fmean(x, g, w, TRA = "-")} is equivalent to \code{TRA(x, fmean(x, g, w), "-", g)}.
Additionally, a \code{set} argument can be passed to \emph{Fast Statistical Functions} to toggle transformation by reference. For example \code{fmean(iris$Sepal.Length, g = iris$Species, TRA = "fill", set = TRUE)} would modify \code{Sepal.Length} in-place and return it invisibly. \newline

Having grouping and data transformation functionality directly built into generic statistical functions facilitates and speeds up many common operations. It notably avoids the need to convert atomic objects to data frames for grouped aggregations or transformations. The \href{https://sebkrantz.github.io/collapse/reference/fast-statistical-functions.html}{\emph{Fast Statistical Functions}} are complemented by a smaller set of \href{https://sebkrantz.github.io/collapse/reference/data-transformations.html}{\emph{Data Transformation Functions}}, including \fct{TRA}, infix functions such as \code{\%r+\%}, \code{\%c+\%}, \code{\%+=\%} and \fct{setop} for row/column-wise arithmetic operations (by reference), \fct{dapply} to apply functions across rows or columns of matrices/data frames, \fct{BY} for general split-apply-combine computing, and specialized functions such as \fct{fscale} or \fct{f[hd]within} for (grouped, weighted) scaling and centering:
\begin{Schunk}
\begin{Sinput}
R> BY(num_vars(iris), g = iris$Species, FUN = mad) |> head(1)
\end{Sinput}
\begin{Soutput}
       Sepal.Length Sepal.Width Petal.Length Petal.Width
setosa       0.2965      0.3706       0.1483           0
\end{Soutput}
\end{Schunk}
\begin{Schunk}
\begin{Sinput}
R> fscale(iris$Sepal.Length, g = iris$Species) |> fsd(g = iris$Species)
\end{Sinput}
\begin{Soutput}
    setosa versicolor  virginica
         1          1          1
\end{Soutput}
\end{Schunk}
\section{Integration with data manipulation functions} \label{sec:integration}
\pkg{collapse} also provides a broad set of \href{https://sebkrantz.github.io/collapse/reference/fast-data-manipulation.html}{Fast Data Manipulation Functions}, including \fct{fselect}, \fct{fsubset}, \fct{fslice}, \fct{fgroup\_by}, \fct{fsummarise}, \fct{fmutate}, \fct{frename}, \fct{fcount}, etc. as optimized analogues to base \proglang{R}/\pkg{tidyverse} functions. These are integrated with the \emph{Fast Statistical and Transformation Functions} to enable vectorized operations in a familiar data frame oriented and \pkg{tidyverse}-like workflow. I illustrate this using the included World Development Dataset (\href{https://sebkrantz.github.io/collapse/reference/wlddev.html}{\code{wlddev}}) recording five key indicators for 216 economies and years 1960-2020.
\begin{Schunk}
\begin{Sinput}
R> fndistinct(wlddev)
\end{Sinput}
\begin{Soutput}
country   iso3c    date    year  decade  region  income    OECD   PCGDP
    216     216      61      61       7       7       4       2    9470
 LIFEEX    GINI     ODA     POP
  10548     368    7832   12877
\end{Soutput}
\end{Schunk}
Below, I track changes in Life Expectancy at the country level, computing the range, average change per year, and the correlation with GDP per Capita in log-differences\footnote{\fct{Dlog} abbreviates \code{fdiff(log(x))}, \fct{pwcor} wraps \code{cor(..., use = "pairwise.complete.obs")}.} while also saving the latest measurements.\footnote{Note that \fct{fsummarise} never re-groups data, making a call to \fct{fungroup} redundant.} The \fct{collap} function called at the end to aggregate the country-level statistics across income groups is a convenience function for mixed-type aggregation. By default, it uses \fct{fmean} for numeric data and \fct{fmode} for categorical data, here also applying population weights such that the most populous country is selected.\footnote{See also \href{https://sebkrantz.github.io/Rblog/2021/01/09/advanced-data-aggregation/}{this blog post on aggregating survey data using \pkg{collapse}} which showcases more aspects of the \fct{collap} function using real census data.}
\begin{Schunk}
\begin{Sinput}
R> wlddev |> fgroup_by(country, income) |>
+    fsummarise(n = fnobs(LIFEEX),
+               diff = fmax(LIFEEX) %-=% fmin(LIFEEX),
+               mean_diff = fmean(fdiff(LIFEEX)),
+               cor_PCGDP = pwcor(Dlog(LIFEEX), Dlog(PCGDP)),
+               across(c(LIFEEX, PCGDP, POP), flast)) |>
+    collap( ~ income, w = ~ POP, keep.w = FALSE) |> print(digits = 2)
\end{Sinput}
\begin{Soutput}
        country              income  n diff mean_diff cor_PCGDP LIFEEX PCGDP
1 United States         High income 60   13      0.23  -0.00828     81 44599
2      Ethiopia          Low income 60   26      0.43   0.27153     64   691
3         India Lower middle income 60   24      0.41  -0.15366     69  2324
4         China Upper middle income 60   27      0.45   0.00077     76  8749
\end{Soutput}
\end{Schunk}
With the exception of \code{pwcor(Dlog(LIFEEX), Dlog(PCGDP))}, which is evaluated for every country, all other expressions in \fct{fsummarise} and also the final \fct{collap} call are fully vectorized, i.e., \emph{Fast Statistical and Transformation Functions} are evaluated only once with grouping information passed to their \code{g} arguments. Using subtraction by reference in the range is more memory efficient as the executed expression is equivalent to \code{fmax(LIFFEX, country) \%-=\% fmin(LIFFEX, country)}. \fct{across} directly invokes \fct{flast.data.frame} on the subset of columns. \emph{Fast Statistical Functions} also have a method for grouped data, thus \fct{fsummarise} is not always needed. Below, I average recent data with population weights.\footnote{Functions like \fct{fmean}, when called on a grouped data frame with weights, by default (\code{keep.w = TRUE}) sum the weights column and retain it after the grouping columns to permit further weighted operations. \vspace{-10mm}}
\begin{Schunk}
\begin{Sinput}
R> wlddev |> fsubset(year >= 2015, income, PCGDP:POP) |>
+    fgroup_by(income) |> fmean(POP)
\end{Sinput}
\begin{Soutput}
               income   sum.POP PCGDP LIFEEX  GINI        ODA
1         High income 5.901e+09 43340  80.70 36.14   81398948
2          Low income 3.296e+09   663  63.05 39.13 2466732035
3 Lower middle income 1.490e+10  2177  68.31 36.48 2234540565
4 Upper middle income 1.318e+10  8168  75.51 41.68  -86630117
\end{Soutput}
\end{Schunk}
Emplyoing the \emph{Fast Statistical Functions} directly often results in more efficient code. For example, computing population shares for each year as \code{fmutate(wlddev, POP\_share = fsum\\(POP, year, TRA = "/"))} is considerably more efficient than \code{wlddev |> fgroup_by(year) |> fmutate(POP\_share = fsum(POP, TRA = "/")) |> fungroup()}, which in turn is more efficient than using \code{POP / fsum(POP)} (extra full-length allocation) or \code{proportions(POP)} (slow split-apply-combine evaluation logic) inside the grouped \fct{fmutate} call. \newline

\pkg{collapse} thus offers several ways to reach the same outcome. This may be confusing at first, but all options can be distinguished in terms of efficiency. Hence, it is helpful to think about computer resources and attempt to craft minimalistic solutions to complex problems. For example, I recently combined multiple spatial datasets on points of interest (POIs) and needed to deduplicate them. I decided to keep the richest source for each location and POI type/category. After creating comparable POI confidence, location, and type indicators, my deduplication expression ended up being a single line of the form \code{fsubset(data, source == fmode(source, list(location, type), confidence, "fill"))}---which retains POIs from the confidence-weighted most frequent (richest) source by location and type. This is efficient because it avoids materializing intermediate datasets and relegates all computations to \fct{fmode}---\emph{Fast Statistical Functions} call a highly optimized internal grouping algorithm.
\subsection{Vectorizations for advanced tasks} \label{ssec:vfat}
\fct{fsummarise} and \fct{fmutate} follow an eager vectorization approach, such that using any \emph{Fast Statistical Function} in an expression causes the entire expression to be vectorized (i.e., evaluated only once with \emph{Fast Statistical Function} returning grouped output). This only applies to visible expressions, \pkg{collapse} cannot read the contents of custom functions, thus such functions are always evaluated using split-apply-combine logic. This eager vectorization approach enables efficient grouped calculations of more complex statistics. Below, I forecast the population of each region via linear regression (\code{POP ~ year}) in a fully vectorized way.
\begin{Schunk}
\begin{Sinput}
R> wlddev |> collap(POP ~ region + year, FUN = fsum) |>
+    fmutate(POP = POP / 1e6) |> fgroup_by(region) |>
+    fmutate(dmy = fmean(year, TRA = "-")) |>
+    fsummarise(beta = fsum(POP, dmy) %/=% fsum(dmy, dmy),
+               POP20 = flast(POP)) |>
+    fmutate(POP21 = POP20 + beta, POP22 = POP21 + beta,
+            POP23 = POP22 + beta, POP24 = POP23 + beta)
\end{Sinput}
\begin{Soutput}
                      region   beta  POP20  POP21  POP22  POP23  POP24
1        East Asia & Pacific 18.731 2291.4 2310.2 2328.9 2347.6 2366.4
2      Europe & Central Asia  2.463  921.2  923.7  926.1  928.6  931.0
3  Latin America & Caribbean  6.460  646.4  652.9  659.4  665.8  672.3
4 Middle East & North Africa  5.409  456.7  462.1  467.5  472.9  478.3
5              North America  2.301  365.9  368.2  370.5  372.8  375.1
6                 South Asia 19.640 1835.8 1855.4 1875.1 1894.7 1914.3
7         Sub-Saharan Africa 12.852 1103.5 1116.3 1129.2 1142.0 1154.9
\end{Soutput}
\end{Schunk}
When \fct{fsummarise} evaluates an expression involving \emph{Fast Statistical Functions}, it sets their \code{g} argument with a grouping (\class{GRP}) object that is directly handed to \proglang{C/C++}, and also sets \code{use.g.names = FALSE}. Hence, weights (\code{w}) becomes the second positional argument. Similarly, \fct{fmutate} sets \code{g} and \code{TRA = "fill"}, which can be overwritten by the user (here with \code{TRA = "-"}). The expression \code{fsum(x, dmy) \%/=\% fsum(dmy, dmy)} amounts to \code{cov(x, y)/var(y)}, but is vectorized across groups and memory efficient---leveraging the weights (\code{w}) argument to \fct{fsum} to compute products (\code{v * dmy} and \code{dmy * dmy}) internally and division by reference (\code{\%/=\%}) to avoid an additional allocation. In a 2023 \href{https://sebkrantz.github.io/Rblog/2023/04/12/collapse-and-the-fastverse-reflecting-the-past-present-and-future/}{blog post}, I forecasted high-resolution population estimates for South Africa like this. Using 1 $km^2$ \href{https://www.worldpop.org/}{WorldPop} data available for years 2014-2020, I ran 1.6 million cell-regressions and obtained 2 forecasts for 2021 and 2022 in less than 0.3 seconds on my M1 Mac. Another neat example from the community, shared by Andrew Ghazi in a \href{https://andrewghazi.github.io/posts/collapse\_is\_sick/sick.html}{blog post}, vectorizes an expression to compute the $p$~value, \code{2 * pt(abs(fmean(x) * sqrt(6) / fsd(x)), 5, lower.tail = FALSE)}, across 300,000 groups for a simulation study, yielding a 70x performance increase over \pkg{dplyr}. \newline
% The eager vectorization approach of \pkg{collapse} here replaces \code{fmean(x)} and \code{fsd(x)} by their grouped versions and evaluates the entire expression once rather than 300k times as in \pkg{dplyr}. \newline

\pkg{collapse} also vectorizes advanced statistics. The following calculates a weighted set of summary statistics by groups, with weighted quantiles type 8 following \citet{hyndman1996sample}.\footnote{\pkg{collapse} computes weighted quantiles in a theoretically consistent way, see \href{https://sebkrantz.github.io/collapse/reference/fquantile.html}{fquantile} for details.}
\begin{Schunk}
\begin{Sinput}
R> wlddev |> fsubset(is.finite(POP)) |> fgroup_by(region) |>
+    fmutate(o = radixorder(GRPid(), LIFEEX)) |>
+    fsummarise(min = fmin(LIFEEX),
+               Q1 = fnth(LIFEEX, 0.25, POP, o = o, ties = "q8"),
+               mean = fmean(LIFEEX, POP),
+               median = fmedian(LIFEEX, POP, o = o),
+               Q3 = fnth(LIFEEX, 0.75, POP, o = o, ties = "q8"),
+               max = fmax(LIFEEX))
\end{Sinput}
\begin{Soutput}
                      region   min    Q1  mean median    Q3   max
1        East Asia & Pacific 18.91 65.28 68.45  69.67 73.86 85.08
2      Europe & Central Asia 45.37 68.68 72.30  71.58 76.67 85.42
3  Latin America & Caribbean 41.76 65.17 69.16  70.87 74.48 82.19
4 Middle East & North Africa 29.92 61.96 66.65  69.12 72.64 82.80
5              North America 68.90 73.57 75.54  75.62 78.38 82.05
6                 South Asia 32.45 55.08 60.19  62.00 66.67 78.92
7         Sub-Saharan Africa 26.17 46.51 52.53  52.23 58.32 74.51
\end{Soutput}
\end{Schunk}
Weighted quantiles have a sub-column level parallel implementation,\footnote{Use \code{set\_collapse(nthreads = \#)} or set the \code{nthreads} arguments of \fct{fnth}/\fct{fmedian} (default 1).} and, as shown above, can also harness an (optional) optimization via an overall ordering vector---combining groups with the data column to avoid repeated sorting of the same elements by different functions. % \newline
%
% This code is very fast because data does not need to be split by groups. Under the hood it is principally a syntax translation to the low-level API introduced above.\footnote{\fct{fgroup\_by} creates a \class{GRP} object from the \code{income} column, attaches it as an attribute, and \fct{fsummarise}/\fct{across} fetches it and passes it to the \code{g} arguments of the \emph{Fast Statistical Functions} set as a keyword argument (and sets \code{use.g.names = FALSE}).
%
% Thus, \code{w} becomes the second positional argument. Since \fct{fmean} is S3 generic, \fct{across} directly invokes \fct{fmean.data.frame} on the subset of columns. \vspace{-10mm}}  %The following example calculates weighted group means. By default (\code{keep.w = TRUE}) \code{fmean.grouped\_df} also sums the weights in each group.\footnote{\class{grouped\_df} methods in \pkg{collapse} support grouped data created with either \fct{fgroup\_by} or \fct{dplyr::group\_by}. The latter requires an additional \proglang{C} routine to convert the \pkg{dplyr} grouping object to a \class{GRP} object, and is thus less efficient.}
%
\subsection{Grouping objects and lower-level API} \label{ssec:gopt}
Whereas the \code{g} argument supports ad-hoc grouping with vectors and lists/data frames, for repeated operations the cost of grouping can be minimized by using factors (see \code{?qF} for efficient factor generation) or \class{GRP} objects as inputs. The latter contain all information \pkg{collapse}'s statistical functions may require to operate across groups and are thus passed to internal \proglang{C/C++} code without checks. They can be created with \code{GRP()}. Its basic syntax is:
\begin{Code}
GRP(X, by = NULL, sort = TRUE, return.groups = TRUE, method = "auto", ...)
\end{Code}
 Below, I create a \class{GRP} object from two columns in the World Development Dataset (\href{https://sebkrantz.github.io/collapse/reference/wlddev.html}{\code{wlddev}}). The \code{by} argument also supports column names/indices, and \code{X} could also be an atomic vector.
\begin{Schunk}
\begin{Sinput}
R> str(g <- GRP(wlddev, ~ income + OECD))
\end{Sinput}
\begin{Soutput}
Class 'GRP'  hidden list of 9
 $ N.groups    : int 6
 $ group.id    : int [1:13176] 3 3 3 3 3 3 3 3 3 3 ...
 $ group.sizes : int [1:6] 2745 2074 1830 2867 3538 122
 $ groups      :'data.frame':	6 obs. of  2 variables:
  ..$ income: Factor w/ 4 levels "High income",..: 1 1 2 3 4 4
  .. ..- attr(*, "label")= chr "Income Level"
  ..$ OECD  : logi [1:6] FALSE TRUE FALSE FALSE FALSE TRUE
  .. ..- attr(*, "label")= chr "Is OECD Member Country?"
 $ group.vars  : chr [1:2] "income" "OECD"
 $ ordered     : Named logi [1:2] TRUE FALSE
  ..- attr(*, "names")= chr [1:2] "ordered" "sorted"
 $ order       : int [1:13176] 245 246 247 248 249 250 251 252 253 254 ...
  ..- attr(*, "starts")= int [1:6] 1 2746 4820 6650 9517 13055
  ..- attr(*, "maxgrpn")= int 3538
  ..- attr(*, "sorted")= logi FALSE
 $ group.starts: int [1:6] 245 611 1 306 62 7687
 $ call        : language GRP.default(X = wlddev, by = ~income + OECD)
\end{Soutput}
\end{Schunk}
\class{GRP} objects make grouped statistical computations in \pkg{collapse} fully programmable. I can employ the object with the \emph{Fast Statistical Functions} and some utilities\footnote{\fct{add\_vars} is a fast \fct{cbind.data.frame} which also has an assignment method, and \fct{get\_vars} enables fast and secure extraction of data frame columns. \code{use = FALSE} abbreviates \code{use.g.names = FALSE}.} to efficiently aggregate GDP per capita, life expectancy, and country name, again applying population weights.
\begin{Schunk}
\begin{Sinput}
R> add_vars(g$groups,
+   get_vars(wlddev, "country") |> fmode(g, wlddev$POP, use = FALSE),
+   get_vars(wlddev, c("PCGDP", "LIFEEX")) |> fmean(g, wlddev$POP, use = F),
+   get_vars(wlddev, "POP") |> fsum(g, use = FALSE))
\end{Sinput}
\begin{Soutput}
               income  OECD       country   PCGDP LIFEEX       POP
1         High income FALSE  Saudi Arabia 22426.7  73.00 3.114e+09
2         High income  TRUE United States 31749.6  75.84 5.573e+10
3          Low income FALSE      Ethiopia   557.1  53.51 2.095e+10
4 Lower middle income FALSE         India  1238.8  60.59 1.138e+11
5 Upper middle income FALSE         China  3820.6  68.21 1.114e+11
6 Upper middle income  TRUE        Mexico  8311.2  69.06 8.162e+09
\end{Soutput}
\end{Schunk}
The above is equivalent to \code{collap(wlddev, country + PCGDP + LIFEEX ~ income + OECD, w = ~ POP)}, which internally toggles many of the same function calls. \newline

% For advanced data aggregation, \pkg{collapse} also provides a convenience function, \fct{collap}, which, by default, uses \fct{fmean} for numeric, \fct{fmode} for categorical, and \fct{fsum} for weight columns, and preserves their order. The equivalent expression using this function would be
%
% <<collap>>=
% ollap(wlddev, country + PCGDP + LIFEEX ~ income + OECD, w = ~ POP)
% @
%
Similarly, data can be transformed, here using \fct{fwithin} to level average differences in economic status, adding back the overall mean after subtracting out group means:\footnote{\fct{add\_stub} adds a prefix (or suffix if \code{pre = FALSE}) to columns $\to$ \code{center\_PCGDP} and \code{center\_LIFEEX}.}
\begin{Schunk}
\begin{Sinput}
R> add_vars(wlddev) <- get_vars(wlddev, c("PCGDP", "LIFEEX")) |>
+      fwithin(g, mean = "overall.mean") |> add_stub("center_")
\end{Sinput}
\end{Schunk}
%
% For (higher-dimensional) centering, \pkg{collapse} also has specialized function(s) \fct{f[hd]within} with additional options, and \fct{fscale} supports various scaling and centering operations.\newline

The lower-level API is useful for package development and standard-evaluation programming, as further elucidated in the \href{https://sebkrantz.github.io/collapse/articles/developing_with_collapse.html}{vignette on developing with \pkg{collapse}}. Users should note that \pkg{collapse} does not provide metaprogramming capabilities in its non-standard evaluation functions---such as \href{https://rlang.r-lib.org/reference/topic-quosure.html}{quosures} or \href{https://dplyr.tidyverse.org/articles/programming.html}{indirection} familiar to \pkg{tidyverse} users. Instead, it has standard-evaluation equivalents to some of these functions which typically end with a \code{v} for 'variables', such as \fct{collapv}, \fct{fslicev}, \fct{ftransformv}, etc. In other cases, including those typically handled by \fct{fsummarise} or \fct{fmutate}, users need to use the lower-level API for programming, or resort to \fct{substitute} and friends. The main reason for not providing higher-level metaprogramming capabilities is to keep all functions as simple as possible. It also compels users to think deeply about their programs and devise more efficient solutions. \newline

As the small benchmark below illustrates, \pkg{dplyr}'s internally more complex data manipulation functions produce much greater overheads, which can noticeably add-up in longer scripts.\footnote{\fct{bmark} is a slim wrapper around \fct{bench::mark}. See the Computational details section.}
\begin{Schunk}
\begin{Sinput}
R> bmark(collapse = collapse::fsummarise(wlddev, mean = fmean(PCGDP)),
+        dplyr = dplyr::summarise(wlddev, mean = fmean(PCGDP)))
\end{Sinput}
\begin{Soutput}
  expression      min   median mem_alloc n_itr n_gc total_time
1   collapse   6.68µs   8.04µs    1.05KB  9999    1    83.59ms
2      dplyr 215.13µs 254.04µs    1.54MB  9936   64      2.78s
\end{Soutput}
\end{Schunk}
Grouped programming using 'GRP' objects and \emph{Fast Statistical Functions} is also particularly powerful with vectors and matrices. For example, in the \href{https://raw.githubusercontent.com/SebKrantz/collapse/master/misc/useR2022\%20presentation/collapse_useR2022_final.pdf}{useR 2022 presentation}, I aggregate 32 global input-output tables stored as matrices (\code{x}) from the country to the region level using a single grouping object (\code{g}) and expressions like \code{x |> fsum(g) |> t() |> fsum(g) |> t()}---computing 45 million sums crunching 5.7GB of data in $\sim$0.3 seconds on my M1.\footnote{Another recent application with vectors involved numerically optimizing a parameter $a$ in an equation of the form $\sum_i x_{ij}^a\ \forall j\in J$ so as to minimize the deviation from a target $y_j$ where there are $J$ groups (1 million in my case)---see the first example in \href{https://sebkrantz.github.io/Rblog/2023/04/12/collapse-and-the-fastverse-reflecting-the-past-present-and-future/}{this blog post} for an illustration.}
%On an M1 Mac using 4 threads, this computation, involving 44.7 million summations and 2.6 million weighted means, takes only 0.33 seconds.\footnote{Another recent example involved numerically optimizing a parameter $a$ in an equation of the form $y_j = \sum_i x_{ij}^a\ \forall j\in J$ where there are $J$ groups (1 million in my case), and the optimal value of $a$ is determined by the proximity of the aggregated vector \textbf{y} to another vector \textbf{z}. Thus each iteration of the numerical routine raises the vector \textbf{x} to a different power ($a$), sums it in 1 million groups ($j$) to generate \textbf{y}, and computes the Euclidean distance to \textbf{z} (using \code{collapse::fdist}). Without grouping objects and vectorization, this would have been difficult to handle within reasonable computing times (of a few seconds on the M1).}
%
\newpage
\section{Time series and panel series} \label{sec:ts_ps}
\pkg{collapse} also provides a flexible high-performance architecture to perform time aware computations on time series and panel series. In particular, users can either apply time series and panel data transformations by passing individual and/or time identifiers to the respective functions in an ad-hoc fashion, or by using \class{indexed\_frame} and \class{indexes\_series} classes, which implement full and deep indexation for worry-free application in many contexts. Table~\ref{tab:TSfun} compactly summarizes \pkg{collapse}'s architecture for time series and panel data.
\begin{table}[h]
\begin{tabular}{p{\textwidth}}
\emph{Classes, constructors and utilities} \\
\code{findex\_by(), findex(), unindex(), reindex(), timeid(), is\_irregular(), to\_plm()} $+$ S3 methods for \class{indexed\_frame}, \class{indexed\_series} and \class{index\_df} \\\\
\emph{Core time-based functions} \\
\code{flag(), fdiff(), fgrowth(), fcumsum(), psmat()} \\ \code{psacf(), pspacf(), psccf()} \\\\
\emph{Data transformation functions with supporting methods} \\
\code{fscale(), f[hd]between(), f[hd]within()} \\\\
\emph{Data manipulation functions with supporting methods} \\
\code{fsubset(), funique(), roworder[v]()} (internal), \code{na\_omit()} (internal) \\\\
\emph{Summary functions with supporting methods} \\
\code{varying(), qsu()} \\
\end{tabular}
\caption{\label{tab:TSfun} Time series and panel data architecture.}
\end{table}
\subsection{Ad-hoc computations}
Time series functions such as \fct{fgrowth} to compute growth rates are S3 generic and can be applied to most time series classes. In addition to a \code{g} argument for grouping, these functions also have a \code{t} argument for indexation. But first, I provide a basic example of computing the annualized 10-year growth rates in miles flown by airlines in the USA from 1937-1960.
\begin{Schunk}
\begin{Sinput}
R> fgrowth(airmiles, n = 10, power = 1/10) |> na.omit() |> round(1)
\end{Sinput}
\begin{Soutput}
Time Series:
Start = 1947
End = 1960
Frequency = 1
 [1] 31.0 28.7 25.7 22.5 22.5 24.3 24.6 22.6 19.4 14.2 15.3 15.5 15.8 14.3
\end{Soutput}
\end{Schunk}
The results show that the flight volume as been growing steadily but at a decreasing rate. \newline

To illustrate the full capabilities of these time series functions, I generate a sector-level trade dataset of export values (v) by country (c), sector (s), and year (y). Like many detailed trade datasets, it is unbalanced---not all sectors/products are exported by country c in all years.
\begin{Schunk}
\begin{Sinput}
R> set.seed(101)
R> exports <- expand.grid(y = 2001:2010, c = paste0("c", 1:10),
+    s = paste0("s", 1:10)) |> fmutate(v = abs(rnorm(1e3))) |>
+    colorder(c, s) |> fsubset(-sample.int(1e3, 500))
\end{Sinput}
\end{Schunk}
The following extracts one country-sector series from the \code{exports} dataset. It is irregular, missing years 2003 and 2006.\footnote{\code{\%=\%} is an infix operator for the \fct{massign} function in \pkg{collapse} which is a multivariate version of \fct{assign}.} Indexation using the \code{t} argument still allows for correct (time-aware) computations in this context without the need to 'expand' the data/fill gaps.
\begin{Schunk}
\begin{Sinput}
R> .c(y, v) %=% fsubset(exports, c == "c1" & s == "s7", -c, -s)
R> print(y)
\end{Sinput}
\begin{Soutput}
[1] 2001 2002 2004 2005 2007 2008 2009 2010
\end{Soutput}
\begin{Sinput}
R> fgrowth(v, t = y) |> round(2)
\end{Sinput}
\begin{Soutput}
[1]     NA 175.52     NA -22.37     NA 624.27 -79.01 534.56
\end{Soutput}
\begin{Sinput}
R> fgrowth(v, -1:3, t = y) |> head(4)
\end{Sinput}
\begin{Soutput}
        FG1     --     G1   L2G1   L3G1
[1,] -63.71 0.3893     NA     NA     NA
[2,]     NA 1.0726 175.52     NA     NA
[3,]  28.82 0.8450     NA -21.22 117.05
[4,]     NA 0.6559 -22.37     NA -38.85
\end{Soutput}
\end{Schunk}
If \code{t} is a plain numeric vector or factor as in this case, it is coerced to integer and interpreted as time steps.\footnote{This is premised on the observation that the most common form of temporal identifier is a plain numeric vector representing calendar years. Users should manually call \code{timeid()} on plain vectors with decimals.} Time objects like \class{Date} or \class{POSIXct} on the other hand are internally passed through \fct{timeid} to generate an appropriate integer representation of them.\footnote{\fct{timeid} divides by the greatest common divisor (GCD) and subtracts the minimum to generate an integer-id (starting form 1). For this approach to work, \code{t} must have an appropriate class, e.g., for monthly/quarterly data, \code{zoo::yearmon()/zoo::yearqtr()} should be used instead of \class{Date} or \class{POSIXct}.} \newline

Functions \code{flag()/fdiff()/fgrowth()} also have associated 'operators' \code{L()/D()/G()} to facilitate their use inside formulas and provide an enhanced data frame interface for convenient ad-hoc computations. With panel data, \code{t} can be omitted, but this requires sorted data with consecutive groups. %\footnote{This is because a group-lag is computed in a single pass, requiring all group elements to be consecutive.}
Below, I demonstrate two ways to compute a sequence of lagged growth rates using either \code{G()} or \fct{fgrowth} and \fct{tfm}---a shorthand for \fct{ftransform}.\footnote{Several key functions in \pkg{collapse} have syntactic shorthands. The \code{list(v = v)} is needed here to prevent \fct{fgrowth} from creating a matrix with the growth rates, ensuring that the \class{list} method applies.}
\begin{Schunk}
\begin{Sinput}
R> G(exports, -1:2, by = v ~ c + s, t = ~ y) |> head(3)
\end{Sinput}
\begin{Soutput}
   c  s    y  FG1.v      v   G1.v L2G1.v
1 c1 s1 2002 -18.15 0.5525     NA     NA
2 c1 s1 2003 214.87 0.6749  22.17     NA
3 c1 s1 2004 -31.02 0.2144 -68.24  -61.2
\end{Soutput}
\begin{Sinput}
R> tfm(exports, fgrowth(list(v = v), -1:2, g = list(c, s), t = y)) |> head(3)
\end{Sinput}
\begin{Soutput}
   c  s    y      v  FG1.v   G1.v L2G1.v
1 c1 s1 2002 0.5525 -18.15     NA     NA
2 c1 s1 2003 0.6749 214.87  22.17     NA
3 c1 s1 2004 0.2144 -31.02 -68.24  -61.2
\end{Soutput}
\end{Schunk}
These functions and operators are also integrated with \fct{fgroup\_by} and \fct{fmutate} for vectorized computations. As mentioned earlier, ad-hoc grouping is always more efficient.
\begin{Schunk}
\begin{Sinput}
R> A <- exports |> fgroup_by(c, s) |> fmutate(gv = G(v, t = y)) |> fungroup()
R> head(B <- exports |> fmutate(gv = G(v, g = list(c, s), t = y)), 4)
\end{Sinput}
\begin{Soutput}
   c  s    y      v     gv
1 c1 s1 2002 0.5525     NA
2 c1 s1 2003 0.6749  22.17
3 c1 s1 2004 0.2144 -68.24
4 c1 s1 2005 0.3108  44.98
\end{Soutput}
\begin{Sinput}
R> identical(A, B)
\end{Sinput}
\begin{Soutput}
[1] TRUE
\end{Soutput}
\end{Schunk}
%
% Similarly, functions to scale, center, and average data have groups (\code{g}) and also weights (\code{w}) arguments, and corresponding operators \code{STD(),[HD]W(),[HD]B()} to facilitate ad-hoc transformations. Below, two ways to perform grouped scaling are demonstrated. The operator version is slightly faster and renames the transformed columns by default (\code{stub = TRUE}).
% %
% <<example_scale_cener>>=
% iris |> fgroup_by(Species) |> fscale() |> head(2)
% STD(iris, ~ Species) |> head(2)
% @
% The following example demonstrates a fixed-effects regression \`a la \citet{mundlak1978pooling}.
% <<example_reg>>=
% lm(mpg ~ carb + B(carb, cyl), data = mtcars) |> coef()
% @
% \pkg{collapse} also offers higher-dimensional between and within transformations, powered by \proglang{C++} code conditionally imported (and accessed directly) from \pkg{fixest}. The following detrends GDP per Capita and Life Expectancy at Birth using country-specific cubic polynomials.
% <<example_HD>>=
% HDW(wlddev, PCGDP + LIFEEX ~ iso3c * poly(year, 3), stub = F) |> head(2)
% @
%
\subsection{Indexed series and frames}
For more complex use cases, indexation is convenient. \pkg{collapse} supports \pkg{plm}'s \class{pseries} and \class{pdata.frame} classes through dedicated methods. Flexibility and performance considerations lead
to the creation of new classes \class{indexes\_series} and \class{indexed\_frame} which inherit from the former. Any data frame-like object can become an \class{indexed\_frame} and behave as usual for other operations. The technical implementation of these classes is described in the \href{https://sebkrantz.github.io/collapse/articles/collapse_object_handling.html#class-agnostic-grouped-and-indexed-data-frames}{vignette on object handling} and, in more detail, in the \href{https://sebkrantz.github.io/collapse/reference/indexing.html}{documentation}. Their basic usage is:
\begin{Code}
data_ix <- findex_by(data, id1, ..., time)
data_ix$indexed_series; with(data, indexed_series)
index_df <- findex(data_ix)
\end{Code}
Data can be indexed using one or more indexing variables. Unlike \class{pdata.frame}, an \\ \class{indexed\_frame} is a deeply indexed structure---every series inside the frame is already an \class{indexes\_series}. A comprehensive set of \href{https://sebkrantz.github.io/collapse/reference/indexing.html}{methods for subsetting and manipulation}, and applicable \class{pseries} and \class{pdata.frame} methods for time series and transformation functions like \code{flag()/L()} ensure that these objects behave in a time-/panel-aware manner in any caller environment (\fct{with}, \fct{lm}, etc.). % A basic demonstration with World Bank panel data showcases the flexibility of these classes.
Indexation can be undone using \code{unindex()} and redone with \code{reindex()} and a suitable \class{index\_df}. \class{indexes\_series} can be atomic vectors or matrices (including \class{ts} or \class{xts}) and can also be created directly using \code{reindex()}.
\begin{Code}
data <- unindex(data_ix)
data_ix <- reindex(data, index = index_df)
indexed_series <- reindex(vec/mat, index = vec/index_df)
\end{Code}
It is worth highlighting that the flexibility of this architecture is new to the \proglang{R} ecosystem: A \class{pdata.frame} or \class{fixest\_panel} only works inside \pkg{plm}/\pkg{fixest} estimation functions.\footnote{And, in the case of \pkg{fixest}, inside \pkg{data.table} due to dedicated methods.} Time series classes like \class{xts} and \class{tsibble} also do not provide deeply indexed structures or native handling of irregularity in basic operations. \class{indexed\_series} and \class{indexed\_frame}, on the other hand, work 'anywhere', and can be superimposed on any suitable object as long as \pkg{collapse}'s functions (\code{flag()/L()}, etc.) are used to perform the time-based computations. \newline

An example follows using the \code{exports} data. \emph{Note} that data can be unsorted for indexation.
\begin{Schunk}
\begin{Sinput}
R> exportsi <- exports |> findex_by(c, s, y)
R> exportsi |> G(0:1) |> head(5)
\end{Sinput}
\begin{Soutput}
   c  s    y      v   G1.v
1 c1 s1 2002 0.5525     NA
2 c1 s1 2003 0.6749  22.17
3 c1 s1 2004 0.2144 -68.24
4 c1 s1 2005 0.3108  44.98
5 c1 s1 2006 1.1740 277.76

Indexed by:  c.s [1] | y [5 (10)]
\end{Soutput}
\begin{Sinput}
R> exportsi |> findex() |> print(2)
\end{Sinput}
\begin{Soutput}
    c.s    y
1 c1.s1 2002
2 c1.s1 2003
---
499 c10.s10 2007
500 c10.s10 2009

c.s [100] | y [10]
\end{Soutput}
\end{Schunk}
The index statistics are: \code{[N. ids] | [N. periods (total periods: (max-min)/GCD)]}. % This extracts an \class{indexes\_series} and demonstrates its properties.
\begin{Schunk}
\begin{Sinput}
R> vi <- exportsi$v; str(vi, width = 70, strict = "cut")
\end{Sinput}
\begin{Soutput}
 'indexed_series' num [1:500] 0.552 0.675 0.214 0.311 1.174 ...
 - attr(*, "index_df")=Classes 'index_df', 'pindex' and 'data.frame'..
  ..$ c.s: Factor w/ 100 levels "c1.s1","c2.s1",..: 1 1 1 1 1 1 1 1 ..
  ..$ y  : Ord.factor w/ 10 levels "2001"<"2002"<..: 2 3 4 5 6 7 8 9..
  ..- attr(*, "nam")= chr [1:3] "c" "s" "y"
\end{Soutput}
\begin{Sinput}
R> is_irregular(vi)
\end{Sinput}
\begin{Soutput}
[1] TRUE
\end{Soutput}
\begin{Sinput}
R> vi |> psmat() |> head(3)
\end{Sinput}
\begin{Soutput}
      2001  2002  2003  2004  2005 2006  2007   2008  2009  2010
c1.s1   NA 0.552 0.675 0.214 0.311 1.17 0.619 0.1127 0.917 0.223
c2.s1   NA 0.795    NA    NA 0.237   NA    NA 0.0585 0.818    NA
c3.s1   NA 0.709 0.268 1.464    NA   NA 0.467 0.1193 0.467    NA
\end{Soutput}
\begin{Sinput}
R> fdiff(vi) |> psmat() |> head(3)
\end{Sinput}
\begin{Soutput}
      2001 2002   2003   2004   2005  2006   2007   2008  2009   2010
c1.s1   NA   NA  0.122 -0.461 0.0964 0.863 -0.555 -0.506 0.804 -0.694
c2.s1   NA   NA     NA     NA     NA    NA     NA     NA 0.759     NA
c3.s1   NA   NA -0.441  1.196     NA    NA     NA -0.348 0.348     NA
\end{Soutput}
\end{Schunk}
\fct{psmat}, for panel-series to matrix, generates a matrix or array from panel data. Thanks to deep indexation, indexed computations work inside arbitrary data masking environments:
\begin{Schunk}
\begin{Sinput}
R> settransform(exportsi, v_ld = Dlog(v))
R> lm(v_ld ~ L(v_ld, 1:2), exportsi) |> summary() |> coef() |> round(3)
\end{Sinput}
\begin{Soutput}
               Estimate Std. Error t value Pr(>|t|)
(Intercept)      -0.008      0.141  -0.058    0.954
L(v_ld, 1:2)L1   -0.349      0.115  -3.042    0.004
L(v_ld, 1:2)L2   -0.033      0.154  -0.215    0.831
\end{Soutput}
\end{Schunk}
Indexed series/frames also support transformations such as grouped scaling with \fct{fscale} or demeaning with \fct{fwithin}. Functions \fct{psacf}/\fct{pspacf}/\fct{psccf} provide panel-data autocorrelation functions, which are computed using group-scaled and suitably lagged panel-series. The \class{index\_df} attached to these objects can also be used with other general tools such as \fct{collapse::BY} to perform grouped computations using third-party functions: %For example, below I compute a 5-year rolling average ().
\begin{Schunk}
\begin{Sinput}
R> BY(vi, findex(vi)$c.s, data.table::frollmean, 5) |> head(10)
\end{Sinput}
\begin{Soutput}
 [1]     NA     NA     NA     NA 0.5853 0.5986 0.4861 0.6267 0.6092     NA

Indexed by:  c.s [2] | y [9 (10)]
\end{Soutput}
\end{Schunk}
Last but not least, the computational performance of these classes is second to none.\footnote{See, e.g., the small benchmark presented on slide 40 of the \href{https://raw.githubusercontent.com/SebKrantz/collapse/master/misc/useR2022\%20presentation/collapse_useR2022_final.pdf}{useR 2022 presentation}.}
\section{Table joins and pivots} \label{sec:join_pivot}
Among its suite of \href{https://sebkrantz.github.io/collapse/reference/fast-data-manipulation.html}{data manipulation functions}, \pkg{collapse}'s implementations of table \href{https://sebkrantz.github.io/collapse/reference/join.html}{joins} and \href{https://sebkrantz.github.io/collapse/reference/pivot.html}{pivots} are particularly noteworthy since they offer several new features, including rich verbosity for table joins, pivots supporting variable labels, and recast pivots. Both implementations also offer outstanding computational performance, syntax, and memory efficiency.
\subsection{Joins}
Compared to commercial software such as \proglang{STATA} \citep{STATA}, the implementation of joins in most open-source software provides no information on how many records were joined from both tables. This often provokes manual efforts to validate the join operation. \fct{collapse::join} provides many options to understand table join operations. Its syntax is:
\begin{Code}
join(x, y, on = NULL, how = "left", suffix = NULL, validate = "m:m",
  multiple = FALSE, sort = FALSE, keep.col.order = TRUE, verbose = 1,
  drop.dup.cols = FALSE, require = NULL, column = NULL, attr = NULL, ...)
\end{Code}
It defaults to left join and only takes first matches from \code{y} (\code{multiple = FALSE}). Thus, it simply adds columns to \code{x}, which is efficient and sufficient/desired in many cases. By default (\code{verbose = 1}), it prints information about the join operation and number of records joined. \newline

To demonstrate \fct{join}, I generate a small database for a bachelor in economics curriculum. It has a \code{teacher} table of 4 teachers (\code{id}: PK) and a linked (\code{id}: FK) \code{course} table of 5 courses.
\begin{Schunk}
\begin{Sinput}
R> teacher <- data.frame(id = 1:4, names = c("John", "Jane", "Bob", "Carl"),
+    age = c(35, 32, 42, 67), subject = c("Math", "Econ", "Stats", "Trade"))
R> course <- data.frame(id = c(1, 2, 2, 3, 5), semester = c(1, 1, 2, 1, 2),
+    course = c("Math I", "Microecon", "Macroecon", "Stats I", "History"))
R> join(teacher, course, on = "id")
\end{Sinput}
\begin{Soutput}
left join: teacher[id] 3/4 (75%) <1:1st> course[id] 3/5 (60%)
  id names age subject semester    course
1  1  John  35    Math        1    Math I
2  2  Jane  32    Econ        1 Microecon
3  3   Bob  42   Stats        1   Stats I
4  4  Carl  67   Trade       NA      <NA>
\end{Soutput}
\end{Schunk}
Users can request the generation of a \code{.join} column (\code{column = "name"/TRUE}) akin to \proglang{STATA}'s \code{\_merge} column to indicate the origin of records in the joined table---useful with a full join:
\begin{Schunk}
\begin{Sinput}
R> join(teacher, course, how = "full", multiple = TRUE, column = TRUE)
\end{Sinput}
\begin{Soutput}
full join: teacher[id] 3/4 (75%) <1:1.33> course[id] 4/5 (80%)
  id names age subject semester    course   .join
1  1  John  35    Math        1    Math I matched
2  2  Jane  32    Econ        1 Microecon matched
3  2  Jane  32    Econ        2 Macroecon matched
4  3   Bob  42   Stats        1   Stats I matched
5  4  Carl  67   Trade       NA      <NA> teacher
6  5  <NA>  NA    <NA>        2   History  course
\end{Soutput}
\end{Schunk}
An alternative is to request an attribute (\code{attr = "name"/TRUE}) that also summarizes the join operation, including the output of \fct{fmatch}---the workhorse of \fct{join} if \code{sort = FALSE}.
\begin{Code}
R> join(teacher, course, multiple = TRUE, attr = "jn") |> attr("jn") |> str()
\end{Code}
\begin{Schunk}
\begin{Soutput}
left join: teacher[id] 3/4 (75%) <1:1.33> course[id] 4/5 (80%)
List of 3
 $ call   : language join(x = teacher, y = course, multiple = TRUE,"..
 $ on.cols:List of 2
  ..$ x: chr "id"
  ..$ y: chr "id"
 $ match  : 'qG' int [1:5] 1 2 3 4 NA
  ..- attr(*, "N.nomatch")= int 1
  ..- attr(*, "N.groups")= int 5
  ..- attr(*, "N.distinct")= int 4
\end{Soutput}
\end{Schunk}
Users can also invoke the \code{validate} argument to examine the uniqueness of the join keys in either table: passing a '1' produces an error if the respective key is not unique.
\begin{Schunk}
\begin{Sinput}
R> join(teacher, course, on = "id", validate = "1:1") |>
+    tryCatch(error = function(e) strwrap(e) |> cat(sep = "\n"))
\end{Sinput}
\begin{Soutput}
Error in join(teacher, course, on = "id", validate = "1:1"): Join is
not 1:1: teacher (x) is unique on the join columns; course (y) is
not unique on the join columns
\end{Soutput}
\end{Schunk}
Similarly, the \code{require} argument allows users to demand a minimum matching success rate.
\begin{Schunk}
\begin{Sinput}
R> join(teacher, course, on = "id", require = list(x = 0.8)) |>
+    tryCatch(error = function(e) substr(e, 102, 200) |> cat())
\end{Sinput}
\begin{Soutput}
Matched 75.0% of records in table teacher (x), but 80.0% is required
\end{Soutput}
\end{Schunk}
A few further particularities are worth highlighting. First, \fct{join} is class-agnostic and preserves the attributes of \code{x} (any list-based object). It supports 6 different join operations (\code{"left"}, \code{"right"}, \code{"inner"}, \code{"full"}, \code{"semi"}, or \code{"anti"}). This demonstrates the latter two:
\begin{Schunk}
\begin{Sinput}
R> for (h in c("semi", "anti")) join(teacher, course, how = h) |> print()
\end{Sinput}
\begin{Soutput}
semi join: teacher[id] 3/4 (75%) <1:1st> course[id] 3/5 (60%)
  id names age subject
1  1  John  35    Math
2  2  Jane  32    Econ
3  3   Bob  42   Stats
anti join: teacher[id] 3/4 (75%) <1:1st> course[id] 3/5 (60%)
  id names age subject
1  4  Carl  67   Trade
\end{Soutput}
\end{Schunk}
By default (\code{sort = FALSE}), the order of rows in \code{x} is preserved. Setting \code{sort = TRUE} sorts all records in the joined table by the keys.\footnote{This is done using a separate sort-merge-join algorithm, so it is faster than performing a hash join using \fct{fmatch} followed by sorting, particularly if the data is already sorted on the keys. } The join relationship is indicated inside the \code{<>} as the number of records joined from each table divided by the number of unique matches. % In multi-match settings, this will be reflected by few records from \code{y} being used.
\newpage
\fct{join}'s handling of duplicate columns in both tables is also rather special. By default (\code{suffix = NULL}), \fct{join} extracts the name of the \code{y} table and appends \code{y}-columns with it. % \code{x}-columns are not renamed.
\begin{Schunk}
\begin{Sinput}
R> course$names <- teacher$names[course$id]
R> join(teacher, course, on = "id", how = "inner", multiple = TRUE)
\end{Sinput}
\begin{Soutput}
inner join: teacher[id] 3/4 (75%) <1:1.33> course[id] 4/5 (80%)
duplicate columns: names => renamed using suffix '_course' for y
  id names age subject semester    course names_course
1  1  John  35    Math        1    Math I         John
2  2  Jane  32    Econ        1 Microecon         Jane
3  2  Jane  32    Econ        2 Macroecon         Jane
4  3   Bob  42   Stats        1   Stats I          Bob
\end{Soutput}
\end{Schunk}
This is congruent to the principle of adding columns to \code{x} in the default first-match left join by altering this table as little as possible. Alternatively, option \code{drop.dup.cols = "x"/"y"} can be used to remove duplicate columns from either \code{x} or \code{y} before the join operation.
\begin{Schunk}
\begin{Sinput}
R> join(teacher, course, on = "id", multiple = TRUE, drop.dup.cols = "y")
\end{Sinput}
\begin{Soutput}
left join: teacher[id] 3/4 (75%) <1:1.33> course[id] 4/5 (80%)
duplicate columns: names => dropped from y
  id names age subject semester    course
1  1  John  35    Math        1    Math I
2  2  Jane  32    Econ        1 Microecon
3  2  Jane  32    Econ        2 Macroecon
4  3   Bob  42   Stats        1   Stats I
5  4  Carl  67   Trade       NA      <NA>
\end{Soutput}
\end{Schunk}
A final noteworthy feature is that \fct{fmatch} has a built-in overidentification check, which warns if more key columns than necessary to identify the records are provided. This check only triggers with 3+ id columns as for efficiency reasons the first two ids are jointly hashed.

\fct{join} is thus a highly efficient, versatile, and verbose implementation of table joins for \proglang{R}.
\subsection{Pivots}
The reshaping/pivoting functionality of both commercial and open-source software is also presently unsatisfactory for complex datasets such as surveys or disaggregated production, trade, or financial sector data, where variable names resemble codes and variable labels are essential to making sense of the data. Such datasets can presently only be reshaped by losing these labels or manual efforts to retain them. Modern \proglang{R} packages also offer different reshaping functions, such as \fct{data.table::melt}/\fct{tidyr::pivot\_longer} to combine columns and \fct{data.table::dcast}/\fct{tidyr::pivot\_wider} to expand them, requiring users to learn both. Since the depreciation of \pkg{reshape(2)} \citep{rreshape2}, there is also no modern replacement for \code{reshape2::recast()}, requiring \proglang{R} users to consecutively call two functions. \newline

\fct{collapse::pivot} provides a powerful new implementation of reshaping for \proglang{R} addressing these shortcomings. It has a single intuitive syntax to perform 'longer', 'wider', and 'recast' pivots and supports complex labelled data without loss of information. Its basic syntax is: \newpage
\begin{Code}
pivot(data, ids = NULL, values = NULL, names = NULL, labels = NULL,
  how = "longer", na.rm = FALSE, check.dups = FALSE, FUN = "last", ...)
\end{Code}
The demonstration below employs the included Groningen Growth and Development Centre 10-Sector Database (\href{https://sebkrantz.github.io/collapse/reference/GGDC10S.html}{\code{GGDC10S}}) providing long-run internationally comparable data on sectoral productivity performance in Africa, Asia, and Latin America. While the database covers 10 sectors, for the demonstration I only retain Agriculture, Mining, and Manufacturing.\footnote{The \code{"Label"} column is added for demonstration purposes. \fct{namlab} provides a compact overview of variable names and labels stored in \code{attr(column, "label")}, with (optional) additional information/statistics.}
\begin{Schunk}
\begin{Sinput}
R> data <- GGDC10S |>
+   fmutate(Label = ifelse(Variable == "VA", "Value Added", "Employment")) |>
+   fsubset(is.finite(AGR), Country, Variable, Label, Year, AGR:MAN)
R> namlab(data, N = TRUE, Ndistinct = TRUE, class = TRUE)
\end{Sinput}
\begin{Soutput}
  Variable     Class    N Ndist         Label
1  Country character 4364    43       Country
2 Variable character 4364     2      Variable
3    Label character 4364     2          <NA>
4     Year   numeric 4364    67          Year
5      AGR   numeric 4364  4353   Agriculture
6      MIN   numeric 4355  4224        Mining
7      MAN   numeric 4355  4353 Manufacturing
\end{Soutput}
\end{Schunk}
To reshape this dataset into a longer format, it suffices to call \code{pivot(data, ids = 1:4)}. If \code{labels = "name"} is specified, variable labels stored in \code{attr(column, "label")} are saved to an additional column. In addition, \code{names = list(variable = "var_name", value = "val_name")} can be passed to set alternative names for the \code{variable} and \code{value} columns.
\begin{Schunk}
\begin{Sinput}
R> head(dl <- pivot(data, ids = 1:4, names = list("Sectorcode", "Value"),
+    labels = "Sector", how = "longer"))
\end{Sinput}
\begin{Soutput}
  Country Variable       Label Year Sectorcode      Sector Value
1     BWA       VA Value Added 1964        AGR Agriculture 16.30
2     BWA       VA Value Added 1965        AGR Agriculture 15.73
3     BWA       VA Value Added 1966        AGR Agriculture 17.68
4     BWA       VA Value Added 1967        AGR Agriculture 19.15
5     BWA       VA Value Added 1968        AGR Agriculture 21.10
6     BWA       VA Value Added 1969        AGR Agriculture 21.86
\end{Soutput}
\end{Schunk}
\fct{pivot} only requires essential information and intelligently guesses the rest. For example, the same result could have been obtained by specifying \code{values = c("AGR", "MIN", "MAN")} instead of \code{ids = 1:4}. An exact reverse operation can also be specified as \code{pivot(dl, 1:4, "Value", "Sectorcode", "Sector", how = "wider")}, where \code{dl} is the long data. \newline

The second option is a wider pivot with \code{how = "wider"}. Here, \code{names} and \code{labels} can be used to select columns containing the names of new columns and their labels.\footnote{If multiple columns are selected, they are combined using \code{"\_"} for names and \code{" - "} for labels. \vspace{-5mm}} Note below how the labels are combined with existing labels such that also this operation is without loss of information. It is, however, a destructive operation, as with two or more columns selected through \code{values}, \fct{pivot} is not able to reverse it. Further arguments like \code{na.rm}, \code{fill}, \code{drop}, \code{sort}, and \code{transpose} can be invoked to control the casting process/output.
\begin{Schunk}
\begin{Sinput}
R> head(dw <- pivot(data, c("Country", "Year"), names = "Variable",
+    labels = "Label", how = "wider"))
\end{Sinput}
\begin{Soutput}
  Country Year AGR_VA AGR_EMP MIN_VA MIN_EMP MAN_VA MAN_EMP
1     BWA 1964  16.30   152.1  3.494  1.9400 0.7366   2.420
2     BWA 1965  15.73   153.3  2.496  1.3263 1.0182   2.330
3     BWA 1966  17.68   153.9  1.970  1.0022 0.8038   1.282
4     BWA 1967  19.15   155.1  2.299  1.1192 0.9378   1.042
5     BWA 1968  21.10   156.2  1.839  0.7855 0.7503   1.069
6     BWA 1969  21.86   157.4  5.245  2.0314 2.1396   2.124
\end{Soutput}
\begin{Sinput}
R> namlab(dw)
\end{Sinput}
\begin{Soutput}
  Variable                       Label
1  Country                     Country
2     Year                        Year
3   AGR_VA   Agriculture - Value Added
4  AGR_EMP    Agriculture - Employment
5   MIN_VA        Mining - Value Added
6  MIN_EMP         Mining - Employment
7   MAN_VA Manufacturing - Value Added
8  MAN_EMP  Manufacturing - Employment
\end{Soutput}
\end{Schunk}
For the recast pivot (\code{how = "recast"}), unless a column named \code{variable} exists in the data, the source and (optionally) destination of variable names needs to be specified using a list passed to \code{names}, and similarly for \code{labels}. Again, taking along labels is entirely optional---omitting either the labels-list's \code{from} or \code{to} elements will omit the respective operation.
\begin{Schunk}
\begin{Sinput}
R> head(dr <- pivot(data, c("Country", "Year"),
+    names = list(from = "Variable", to = "Sectorcode"),
+    labels = list(from = "Label", to = "Sector"), how = "recast"))
\end{Sinput}
\begin{Soutput}
  Country Year Sectorcode      Sector    VA   EMP
1     BWA 1964        AGR Agriculture 16.30 152.1
2     BWA 1965        AGR Agriculture 15.73 153.3
3     BWA 1966        AGR Agriculture 17.68 153.9
4     BWA 1967        AGR Agriculture 19.15 155.1
5     BWA 1968        AGR Agriculture 21.10 156.2
6     BWA 1969        AGR Agriculture 21.86 157.4
\end{Soutput}
\begin{Sinput}
R> vlabels(dr)[3:6]
\end{Sinput}
\begin{Soutput}
   Sectorcode        Sector            VA           EMP
           NA            NA "Value Added"  "Employment"
\end{Soutput}
\end{Schunk}
This (\code{dr}) is the tidy format \citep{rtidydata} where each variable is a separate column. It is analytically more useful, e.g., to compute labor productivity as \code{settransform(dr, LP = VA / EMP)} or to estimate a panel-regression with sector fixed-effects. The recast pivot is thus a natural operation to change data representation. As with the other pivots, it preserves all information and can be reversed by simply swapping the contents of the \code{from} and \code{to} keywords. \newline

\fct{pivot} also supports fast aggregation pivots, the default being \code{FUN = "last"}, which simply overwrites values in appearance order if the combination of \code{ids} and \code{names} does not fully identify the data. The latter can be checked with \code{check.dups = TRUE}. A small number extremely fast internal aggregation functions: \code{"first", "last", "sum", "mean", "min", "max",} and \code{"count"}, operate 'on the fly' during reshaping. \fct{pivot} also supports \emph{Fast Statistical Functions}, which will yield vectorized aggregations, but requires a deep copy of the columns aggregated which is avoided by the internal functions. The following example performs aggregation across years with the internal mean function during a recast pivot.
\begin{Schunk}
\begin{Sinput}
R> head(dr_agg <- pivot(data, "Country", c("AGR", "MIN", "MAN"),
+    how = "recast", names = list(from = "Variable", to = "Sectorcode"),
+    labels = list(from = "Label", to = "Sector"), FUN = "mean"))
\end{Sinput}
\begin{Soutput}
  Country Sectorcode      Sector       VA      EMP
1     BWA        AGR Agriculture    462.2   188.06
2     ETH        AGR Agriculture  34389.9 17624.34
3     GHA        AGR Agriculture   1549.4  3016.04
4     KEN        AGR Agriculture 139705.9  5348.91
5     MWI        AGR Agriculture  28512.6  2762.62
6     MUS        AGR Agriculture   3819.6    59.34
\end{Soutput}
\end{Schunk}
The \href{https://sebkrantz.github.io/collapse/reference/pivot.html#ref-examples}{documentation examples} demonstrate more features of \fct{pivot}. Notably, it can also perform longer and recast pivots without id variables, similar to \code{data.table::transpose()}.
\section{List processing} \label{sec:list_proc}
Often in programming, nested structures are needed. A typical use case involves running statistical procedures for multiple configurations of variables and parameters and saving multiple objects, such as a model predictions and performance statistics, in a list. Nested data is also often the result of web scraping or web APIs. A typical use case in development involves serving different data according to user choices. Except for certain recursive functions in packages such as \pkg{purr}, \pkg{tidyr}, or \pkg{rrapply}, \proglang{R} lacks a general recursive toolkit to create, query, and tidy nested data. \pkg{collapse}'s \href{https://sebkrantz.github.io/collapse/reference/list-processing.html}{list processing functions} attempt to provide a basic toolkit. \newline

To create nested data, \fct{rsplit} generalizes \fct{split} and (recursively) splits up data frame-like objects into (nested) lists. For example, we can split the \code{GGDC10S} data by country and variable, such that agricultural employment in Argentina can be accessed as:\footnote{If a nested structure is not needed, \code{flatten = TRUE} lets \fct{rsplit} operate like a faster version of \fct{split}.} \newpage
\begin{Schunk}
\begin{Sinput}
R> d_list <- GGDC10S |> rsplit( ~ Country + Variable)
R> d_list$ARG$EMP$AGR[1:12]
\end{Sinput}
\begin{Soutput}
 [1] 1800 1835 1731 2030 1889 1843 1789 1724 1678 1725 1650 1553
\end{Soutput}
\end{Schunk}
This is a convenient data representation for \emph{Shiny Apps} where we can let the user choose data (e.g., \code{d_list[[input$country]][[input$variable]][[input$sector]]}) without expensive subsetting operations. As mentioned, such data representation can also be the result of an API call parsing JSON or a nested loop or \fct{lapply} call. Below, I write a nested loop running a regression of agriculture on mining and manuacturing output and employment.
\begin{Schunk}
\begin{Sinput}
R> results <- list()
R> for (country in c("ARG", "BRA", "CHL")) {
+    for (variable in c("EMP", "VA")) {
+      m <- lm(log(AGR+1) ~ log(MIN+1) + log(MAN+1) + Year,
+              data = d_list[[country]][[variable]])
+      results[[country]][[variable]] <- list(model = m, BIC = BIC(m),
+                                             summary = summary(m))
+    }
+  }
\end{Sinput}
\end{Schunk}
This programming may not be ideal for this particular use case as I could have used data frame-based tools and saved the result in a column.\footnote{E.g., \code{GGDC10S |> fgroup\_by(Country, Variable) |> fsummarise(results = my\_fun(lm(log(AGR+1) ~ log(MIN+1) + log(MAN+1) + Year)))} with \code{my\_fun <- function(m) list(list(m, BIC(m), summary(m)))}.}
However, there are limits to such workflows. For example, I recently trained a complex ML model for different variables and parameters, while also loading a different dataset for each combination. Loops are useful in such cases, and lists a natural vehicle to structure complex outputs. The main issue with nested lists is that they are complex to query. What if we wanted to know the $R^2$ of these 6 models? We would need to use, e.g., \code{results$ARG$EMP$summary$r.squared} for each model.\newline

This nested list-access problem was the main reason for creating \fct{get\_elem}: an efficient recursive list-filtering function which, by default, simplifies the list tree as much as possible.
\begin{Schunk}
\begin{Sinput}
R> str(r_sq <- results |> get_elem("r.squared"))
\end{Sinput}
\begin{Soutput}
List of 3
 $ ARG:List of 2
  ..$ EMP: num 0.907
  ..$ VA : num 1
 $ BRA:List of 2
  ..$ EMP: num 0.789
  ..$ VA : num 0.999
 $ CHL:List of 2
  ..$ EMP: num 0.106
  ..$ VA : num 0.999
\end{Soutput}
\begin{Sinput}
R> rowbind(r_sq, idcol = "Country", return = "data.frame")
\end{Sinput}
\begin{Soutput}
  Country    EMP     VA
1     ARG 0.9068 0.9996
2     BRA 0.7888 0.9988
3     CHL 0.1058 0.9991
\end{Soutput}
\end{Schunk}
%
% Thus \code{get\_elem("R2")} returns the same nested list with the $R^2$ in all final nodes. Then \code{rowbind()} could be used to create a data frame:
Note how the \code{"summary"} branch was eliminated since it is common to all final nodes; \code{results |> get\_elem("r.squared", keep.tree = TRUE)} could have been used to keep it. \fct{rowbind} then efficiently combines lists of lists. We can also apply \fct{t\_list} to turn the list inside-out:
\begin{Schunk}
\begin{Sinput}
R> r_sq |> t_list() |> rowbind(idcol = "Variable", return = "data.frame")
\end{Sinput}
\begin{Soutput}
  Variable    ARG    BRA    CHL
1      EMP 0.9068 0.7888 0.1058
2       VA 0.9996 0.9988 0.9991
\end{Soutput}
\end{Schunk}
\fct{rowbind} is limited if \fct{get\_elem} returns a more nested or asymmetric list, potentially with vectors/arrays in the final nodes. Suppose we wanted to extract the coefficient matrices:
\begin{Schunk}
\begin{Sinput}
R> results$ARG$EMP$summary$coefficients
\end{Sinput}
\begin{Soutput}
              Estimate Std. Error  t value  Pr(>|t|)
(Intercept)  26.583617  1.2832583  20.7157 1.747e-28
log(MIN + 1)  0.083168  0.0352493   2.3594 2.169e-02
log(MAN + 1) -0.064413  0.0767614  -0.8391 4.048e-01
Year         -0.009683  0.0005556 -17.4278 1.003e-24
\end{Soutput}
\end{Schunk}
For such cases, \fct{unlist2d} provides a complete recursive generalization of \fct{unlist}. It creates a \class{data.frame} or \class{data.table} representation of any nested list using recursive row-binding and coercion operations, while generating (optional) id variables representing the list tree and (optionally) saving row names of matrices or data frames. In the present example
\begin{Schunk}
\begin{Sinput}
R> results |> get_elem("coefficients") |> get_elem(is.matrix) |>
+    unlist2d(idcols = c("Country", "Variable"),
+             row.names = "Covariate") |> head(3)
\end{Sinput}
\begin{Soutput}
  Country Variable    Covariate Estimate Std. Error t value  Pr(>|t|)
1     ARG      EMP  (Intercept) 26.58362    1.28326 20.7157 1.747e-28
2     ARG      EMP log(MIN + 1)  0.08317    0.03525  2.3594 2.169e-02
3     ARG      EMP log(MAN + 1) -0.06441    0.07676 -0.8391 4.048e-01
\end{Soutput}
\end{Schunk}
where \code{get\_elem(is.matrix)} is needed because the models also contain \code{"coefficients"}. \newline

This exemplifies the power of these tools to create, query, and combine nested data in very general ways.
  % I use it extensively to fetch and compare different statistics from (sometimes large) nested lists with different statistical model outputs.
Further useful functions include \fct{has\_elem} to check for the existence of elements, \fct{ldepth} to return the maximum level of recursion, \fct{is\_unlistable} to check whether a list has atomic elements in all final nodes, \fct{[ir]reg\_elem} to recursively extract the (non-)atomic elements, and \fct{rapply2d} to apply functions to nested lists of data objects.
\section{Summary statistics} \label{sec:summ_stat}
\pkg{collapse}'s \href{https://sebkrantz.github.io/collapse/reference/summary-statistics.html}{summary statistics functions} offer a parsimonious yet powerful toolkit to examine complex datasets. A particular focus has been on providing tools to understand multilevel (panel) data. Recall the World Development panel dataset (\href{https://sebkrantz.github.io/collapse/reference/wlddev.html}{\code{wlddev}}) from Section~\ref{sec:integration}. The function \fct{varying} can be used to examine which of the variables are time-varying.
\begin{Schunk}
\begin{Sinput}
R> varying(wlddev, ~ iso3c)
\end{Sinput}
\begin{Soutput}
      country          date          year        decade        region
        FALSE          TRUE          TRUE          TRUE         FALSE
       income          OECD         PCGDP        LIFEEX          GINI
        FALSE         FALSE          TRUE          TRUE          TRUE
          ODA           POP  center_PCGDP center_LIFEEX
         TRUE          TRUE          TRUE          TRUE
\end{Soutput}
\end{Schunk}
A related exercise is to decompose the variance of a panel series into variation between countries and variation within countries over time. Using the (de-)meaning functions \fct{fbetween}/ \fct{fwithin} supporting \class{indexed\_series} (Table~\ref{tab:TSfun}), this is easily demonstrated:
\begin{Schunk}
\begin{Sinput}
R> LIFEEXi <- reindex(wlddev$LIFEEX, index = wlddev$iso3c)
R> all.equal(fvar(LIFEEXi), fvar(fbetween(LIFEEXi)) + fvar(fwithin(LIFEEXi)))
\end{Sinput}
\begin{Soutput}
[1] TRUE
\end{Soutput}
\end{Schunk}
The \fct{qsu} (quick-summary) function provides an efficient method to compute this decomposition, considering the group-means instead of the between transformation\footnote{This is more efficient and equal to using the between transformation if the panel is balanced.} and adding the overall mean back to the within transformation to preserve the scale of the data.
\begin{Schunk}
\begin{Sinput}
R> qsu(LIFEEXi)
\end{Sinput}
\begin{Soutput}
             N/T     Mean       SD      Min      Max
Overall    11670  64.2963  11.4764   18.907  85.4171
Between      207  64.9537   9.8936  40.9663  85.4171
Within   56.3768  64.2963   6.0842  32.9068  84.4198
\end{Soutput}
\end{Schunk}
The decomposition above implies more variation in life expectancy between countries than within countries over time. It can also be computed for different subgroups and with sampling weights. \fct{qsu} also has a data frame method, and by default computes simple statistics.\footnote{The \code{pid} argument to \fct{qsu} can also be used to manually pass identifiers for panel-decomposition, e.g., \code{pid = iso3c}. With indexed data, it is automatically set to the first column in the index (\code{effect = 1}).} Below, I take the latest post-2015 estimates and summarise LIFEEX by income groups with population weights. The \code{WeightSum} column thus records the total population in each group.
\begin{Schunk}
\begin{Sinput}
R> wlda15 <- wlddev |> fsubset(year >= 2015) |> fgroup_by(iso3c) |> flast()
R> qsu(wlda15, by = LIFEEX ~ income, w = ~ POP)
\end{Sinput}
\begin{Soutput}
                      N       WeightSum     Mean      SD      Min     Max
High income          68  1.19122607e+09   80.879   2.441  70.6224  85.078
Low income           29      694'893773  63.8061  3.9266   53.283  72.697
Lower middle income  47  3.06353648e+09  68.7599  4.7055   54.331  76.699
Upper middle income  55  2.67050662e+09  75.9476  2.3895   58.735  80.279
\end{Soutput}
\end{Schunk}
%
% The output shows that the variation in life expectancy is significantly larger for non-OECD countries. For the latter the between- and within-country variation is approximately equal.\footnote{\fct{qsu} also has a convenient formula interface to perform these transformations in an ad-hoc fashion, e.g., the above can be obtained using \code{qsu(wlddev, LIFEEX $\sim$ OECD, $\sim$ iso3c, $\sim$ POP)}, without prior indexation.}
For greater detail, \fct{descr} provides a rich (grouped, weighted) statistical description. It does not support panel-variance decompositions like \fct{qsu}, but also computes detailed frequency tables for categorical data. Below, I summarize income and LIFEEX by OECD membership, scaling the weights to limit printout and replacing missing weights with \code{0} (the default).\footnote{This is necessary in \fct{descr} because \fct{fquantile} does not support missing weights for non-missing \code{x}.\vspace{-5mm}}
\begin{Schunk}
\begin{Sinput}
R> descr(wlda15, by = income + LIFEEX ~ OECD, w = ~ replace_na(POP / 1e6))
\end{Sinput}
\begin{Soutput}
Dataset: wlda15, 2 Variables, N = 216, WeightSum = 7620.902563
Grouped by: OECD [2]
         N   Perc  WeightSum   Perc
FALSE  180  83.33    6311.15  82.81
TRUE    36  16.67    1309.75  17.19
-----------------------------------------------------------------------------
income (factor): Income Level
Statistics (WeightSum = 7621, 0% NAs)
       WeightSum   Perc  Ndist
FALSE    6311.15  82.81      4
TRUE     1309.75  17.19      2

Table (WeightSum Perc)
                         FALSE       TRUE      Total
Lower middle income  3064 48.5     0  0.0  3064 40.2
Upper middle income  2460 39.0   211 16.1  2671 35.0
High income            93  1.5  1099 83.9  1192 15.6
Low income            695 11.0     0  0.0   695  9.1
-----------------------------------------------------------------------------
LIFEEX (numeric): Life expectancy at birth, total (years)
Statistics (N = 199, 7.87% NAs)
         N  Ndist  WeightSum   Perc   Mean    SD    Min    Max   Skew  Kurt
FALSE  163    164    6310.41  82.81  71.14  5.77  53.28  85.08  -0.99  3.76
TRUE    36     36    1309.75  17.19  80.32  2.77  75.05  84.36  -0.29  2.11

Quantiles
          1%     5%    10%    25%    50%    75%    90%    95%    99%
FALSE  54.69  59.38  63.67  69.64  71.71  76.89  76.91  76.91  77.94
TRUE   75.07  75.14  76.03  78.77  80.86  82.86  83.61  84.05   84.3
-----------------------------------------------------------------------------
\end{Soutput}
\end{Schunk}

\fct{descr} also has a \code{stepwise} argument to describe one variable at a time, allowing users to naturally 'step-through' the variables in a large dataset while spreading the computational burden. The \href{https://sebkrantz.github.io/collapse/reference/descr.html}{documentation} provides more details and examples. Both \fct{qsu} and \fct{descr} come with an \code{as.data.frame()} method for efficient tidying and easy further analysis. \newline

A final noteworthy function from \pkg{collapse}'s descriptive statistics toolkit is \fct{qtab}, an enhanced drop-in replacement for \fct{table}. It is enhanced both in a statistical and a computational sense, providing a remarkable performance boost, an option \code{sort = FALSE} to preserve the first-appearance-order of vectors being cross-tabulated, support for frequency weights (\code{w}), and the ability to compute different statistics representing table entries using these weights---vectorized when using \emph{Fast Statistical Functions} as demonstrated below.
\begin{Schunk}
\begin{Sinput}
R> wlda15 |> with(qtab(OECD, income))
\end{Sinput}
\begin{Soutput}
       income
OECD    High income Low income Lower middle income Upper middle income
  FALSE          45         30                  47                  58
  TRUE           34          0                   0                   2
\end{Soutput}
\end{Schunk}
This shows the total population (latest post-2015 estimates) in millions.
\begin{Schunk}
\begin{Sinput}
R> wlda15 |> with(qtab(OECD, income, w = POP / 1e6))
\end{Sinput}
\begin{Soutput}
       income
OECD    High income Low income Lower middle income Upper middle income
  FALSE       93.01     694.89             3063.54             2459.71
  TRUE      1098.75       0.00                0.00              211.01
\end{Soutput}
\end{Schunk}
This shows the average life expectancy in years. The use of \fct{fmean} toggles an efficient vectorized computation of the table entries (i.e., \fct{fmean} is only called once).
\begin{Schunk}
\begin{Sinput}
R> wlda15 |> with(qtab(OECD, income, w = LIFEEX, wFUN = fmean))
\end{Sinput}
\begin{Soutput}
       income
OECD    High income Low income Lower middle income Upper middle income
  FALSE       78.75      62.81               68.30               73.81
  TRUE        81.09                                              76.37
\end{Soutput}
\end{Schunk}
Finally, this calculates a population-weighted average of life expectancy in each group.
\begin{Schunk}
\begin{Sinput}
R> wlda15 |> with(qtab(OECD, income, w = LIFEEX, wFUN = fmean,
+                      wFUN.args = list(w = POP)))
\end{Sinput}
\begin{Soutput}
       income
OECD    High income Low income Lower middle income Upper middle income
  FALSE       77.91      63.81               68.76               75.93
  TRUE        81.13                                              76.10
\end{Soutput}
\end{Schunk}
\class{qtab} objects inherit the \class{table} class, thus all \class{table} methods apply. Apart from the above summary functions, \pkg{collapse} also provides \fct{pwcor}, \fct{pwcov}, and \fct{pwnobs} for convenient (pairwise, weighted) correlations, covariances, and observations counts, respectively.
\section{Global options} \label{sec:glob_opt}
\pkg{collapse} is \href{https://sebkrantz.github.io/collapse/reference/collapse-options.html}{globally configurable} to an extent few packages are: the default values of key function arguments governing the behavior of its algorithms, and the exported namespace, can be adjusted interactively through the \fct{set\_collapse} function. These options are saved in an internal environment called \code{.op}. Its contents can be accessed using \fct{get\_collapse}. \newline

The current set of options comprises the default behavior for missing values (\code{na.rm} arguments in all statistical functions and algorithms), sorted grouping (\code{sort}), multithreading and algorithmic optimizations (\code{nthreads}, \code{stable.algo}), presentational settings (\code{stub}, \code{digits}, \code{verbose}), and, surpassing all else, the package namespace itself (\code{mask}, \code{remove}). \newline

As evident from previous sections, \pkg{collapse} provides performance-improved or otherwise enhanced versions of functions already present in base \proglang{R} (like the \emph{Fast Statistical Functions}, \fct{funique}, \fct{fmatch}, \fct{fsubset}, \fct{ftransform}, etc.) or other packages (especially \pkg{dplyr} \citep{rdplyr}: \fct{fselect}, \fct{fsummarise}, \fct{fmutate}, \fct{frename}, etc.). The objective of being namespace compatible warrants such a naming convention, but this has a syntactical cost, particularly when \pkg{collapse} is used as the primary workhorse package. \newline

To reduce this cost, \pkg{collapse}'s \code{mask} option allows masking existing \proglang{R} functions with the faster \pkg{collapse} versions by creating additional functions in the namespace and instantly exporting them. All \pkg{collapse} functions starting with an 'f' can be passed to the option (with or without the 'f'), e.g., \code{set\_collapse(mask = c("subset", "transform"))} creates \code{subset <- fsubset} and \code{transform <- ftransform} and exports them. Special functions are \code{"n"}, \code{"table"/"qtab"}, and \code{"\%in\%"}, which create \code{n <- GRPN} for use in \code{(f)summarise}/\code{(f)mutate}, \code{table <- qtab}, and replace \code{\%in\%} with a faster version based on \fct{fmatch}, respectively. There also exist several \href{https://sebkrantz.github.io/collapse/reference/collapse-options.html}{convenience keywords to mask related groups of functions}, such as \code{"manip"} (only data manipulation functions), or \code{"all"} (everything), as demonstrated below.
\begin{Code}
set_collapse(mask = "all")
wlddev |> subset(year >= 1990 & is.finite(GINI)) |>
  group_by(year) |>
  summarise(n = n(), across(PCGDP:GINI, mean, w = POP))
with(mtcars, table(cyl, vs, am))
sum(mtcars)
diff(EuStockMarkets)
mean(num_vars(iris), g = iris$Species)
unique(wlddev, cols = c("iso3c", "year"))
range(wlddev$date)
wlddev |> index_by(iso3c, year) |>
  mutate(PCGDP_lag = lag(PCGDP),
         PCGDP_growth = growth(PCGDP)) |> unindex()
\end{Code}
The above is now 100\% \pkg{collapse} code. Similarly, using this option, all code in this article could have been written without f-prefixes. Thus, \pkg{collapse} is able to offer a fast and syntactically clean experience of \proglang{R}---without the need to even restart the session. Masking is completely and interactively reversible: calling \code{set\_collapse(mask = NULL)} instantly removes the additional functions. Option \code{remove} can further be used to remove (un-export) any \pkg{collapse} function, allowing manual conflict management. Function \code{fastverse::fastverse\_conflicts()} from the related \href{https://fastverse.github.io/fastverse/}{\pkg{fastverse} project} \citep{rfastverse} can be used to display namespace conflicts with \pkg{collapse}. Invoking either \code{mask} or \code{remove} detaches \pkg{collapse} and re-attaches it at the top of the search path, letting its namespace to take precedence over other packages. \newline

Such global powers confer responsibilities upon package developers, as further elucidated in the \href{https://sebkrantz.github.io/collapse/articles/developing_with_collapse.html#some-notes-on-global-options}{vignette on developing with \pkg{collapse}}. As a general rule, options \code{mask} and \code{remove} should be off-limits inside packages, and other options need to be reset immediately using \fct{on.exit}.
\section{Benchmark} \label{sec:bench}
This section provides several simple benchmarks to show that \pkg{collapse} provides best-in-\proglang{R} performance for statistics and data manipulation on moderately sized datasets. They are executed on a 2024 Apple MacBook Pro with 48GB M4 Pro chip. It also discusses results from \href{https://github.com/fastverse/fastverse/wiki/Benchmarks}{third-party benchmarks} involving \pkg{collapse}. The first set of benchmarks show that \pkg{collapse} provides faster computationally intensive operations like unique values and matching on large integer and character vectors. It creates integer/character vectors of 10 million obs, with 1000 unique integers and 5776 unique strings, respectively, which are deduplicated/matched in the benchmark. These fast basic operations impact many critical components of the package. % These are serially implemented and critical in supporting more complex functionality. %These algorithms power much of its functionality, such as efficient factor generation with \fct{qF}, cross-tabulation with \fct{qtab}, \fct{join}'s, \fct{pivot}'s, etc.
\begin{Schunk}
\begin{Sinput}
R> set.seed(101);
R> int <- 1:1000; g_int <- sample.int(1000, 1e7, replace = TRUE)
R> char <- c(letters, LETTERS, month.abb, month.name)
R> g_char <- sample(char <- outer(char, char, paste0), 1e7, TRUE)
R> bmark(base = unique(g_int), collapse = funique(g_int))
\end{Sinput}
\begin{Soutput}
  expression     min  median mem_alloc n_itr n_gc total_time
1       base 45.15ms 48.17ms   166.2MB    62   62      3.02s
2   collapse  5.52ms  7.74ms    38.2MB   371   75         3s
\end{Soutput}
\begin{Sinput}
R> bmark(base = unique(g_char), collapse = funique(g_char))
\end{Sinput}
\begin{Soutput}
  expression    min median mem_alloc n_itr n_gc total_time
1       base 71.9ms 73.4ms   166.2MB    41   41      3.02s
2   collapse 14.8ms 17.7ms    38.2MB   165   33      3.01s
\end{Soutput}
\begin{Sinput}
R> bmark(base = match(g_int, int), collapse = fmatch(g_int, int))
\end{Sinput}
\begin{Soutput}
  expression     min  median mem_alloc n_itr n_gc total_time
1       base  17.1ms 17.58ms    76.3MB    79   78       1.4s
2   collapse  7.48ms  8.12ms    38.2MB   260   70      2.11s
\end{Soutput}
\begin{Sinput}
R> bmark(base = match(g_char, char), data.table =
+        chmatch(g_char, char), collapse = fmatch(g_char, char))
\end{Sinput}
\begin{Soutput}
  expression    min median mem_alloc n_itr n_gc total_time
1       base 58.6ms 60.6ms   114.5MB    49   49      3.02s
2 data.table 32.8ms 34.2ms    38.1MB    86   22      3.01s
3   collapse 19.4ms 20.3ms    38.1MB   140   29         3s
\end{Soutput}
\end{Schunk}
The second set below shows that \pkg{collapse}'s statistical functions are very efficient in aggregating a numeric matrix with 10,000 rows and 1000 columns. They are faster than base \proglang{R} even without multithreading, but using 4 threads in this case induces a sizeable difference.
\begin{Schunk}
\begin{Sinput}
R> set_collapse(na.rm = FALSE, sort = FALSE, nthreads = 4)
R> m <- matrix(rnorm(1e7), ncol = 1000)
R> bmark(base = colSums(m), collapse = fsum(m))
\end{Sinput}
\begin{Soutput}
  expression     min   median mem_alloc n_itr n_gc total_time
1       base   5.4ms   6.12ms    7.86KB   492    0      3.01s
2   collapse 292.5µs 387.94µs    7.86KB  7352    1         3s
\end{Soutput}
\begin{Sinput}
R> bmark(base = colMeans(m), collapse = fmean(m))
\end{Sinput}
\begin{Soutput}
  expression      min   median mem_alloc n_itr n_gc total_time
1       base   5.69ms   6.02ms   32.09KB   499    0         3s
2   collapse 285.24µs 393.35µs    7.86KB  7200    0         3s
\end{Soutput}
\begin{Sinput}
R> bmark(matrixStats = matrixStats::colMedians(m), collapse = fmedian(m))
\end{Sinput}
\begin{Soutput}
   expression    min median mem_alloc n_itr n_gc total_time
1 matrixStats 77.1ms   78ms    89.9KB    38    1      2.98s
2    collapse 19.4ms 19.6ms    27.2KB   154    0      3.01s
\end{Soutput}
\end{Schunk}
Below, I also benchmark a grouped version summing the columns within 1000 random groups.
\begin{Schunk}
\begin{Sinput}
R> g <- sample.int(1e3, 1e4, TRUE)
R> bmark(base = rowsum(m, g), collapse = fsum(m, g))
\end{Sinput}
\begin{Soutput}
  expression      min median mem_alloc n_itr n_gc total_time
1       base   5.38ms 5.63ms    7.87MB   466   35      2.63s
2   collapse 918.11µs 1.19ms    7.71MB  1714  127       2.2s
\end{Soutput}
\end{Schunk}
I now turn to basic operations on a medium-sized real-world database recording all flights from New York City (EWR, JFK, and LGA) in 2023---provided by the \pkg{nycflights23} package. The \code{flights} table has 435k flights, and grouping it by day and route yields 76k unique trips.
\begin{Schunk}
\begin{Sinput}
R> fastverse_extend(nycflights23, dplyr, data.table); setDTthreads(4)
R> list(flights, airports, airlines, planes, weather) |> sapply(nrow)
\end{Sinput}
\begin{Soutput}
[1] 435352   1255     14   4840  26207
\end{Soutput}
\begin{Sinput}
R> flights |> fselect(month, day, origin, dest) |> fnunique()
\end{Sinput}
\begin{Soutput}
[1] 75899
\end{Soutput}
\end{Schunk}
In the following, I select 6 numeric variables and sum them across the 76k trips using \pkg{dplyr}, \pkg{data.table}, and \pkg{collapse}. Ostensibly, despite \fct{sum} being 'primitive' (implemented in \proglang{C}), there is a factor 100 between \pkg{dplyr}'s split-apply-combine and \pkg{collapse}'s fully vectorized execution.
\begin{Schunk}
\begin{Sinput}
R> vars <- .c(dep_delay, arr_delay, air_time, distance, hour, minute)
R> bmark(dplyr = flights |> group_by(month, day, origin, dest) |>
+                  summarise(across(all_of(vars), sum), .groups = "drop"),
+        data.table = qDT(flights)[, lapply(.SD, sum), .SDcols = vars,
+                                  by = .(month, day, origin, dest)],
+        collapse = flights |> fgroup_by(month, day, origin, dest) |>
+                     get_vars(vars) |> fsum())
\end{Sinput}
\begin{Soutput}
  expression      min   median mem_alloc n_itr n_gc total_time
1      dplyr 356.33ms 490.66ms    50.5MB     7   69      3.24s
2 data.table   7.37ms    8.5ms    20.8MB   315   33         3s
3   collapse   3.21ms   3.78ms     9.2MB   696   27         3s
\end{Soutput}
\end{Schunk}
Below, I also benchmark the mean and median functions in the same way. It is evident that with non-primitive \proglang{R} functions the split-apply-combine logic becomes even more costly.
\begin{Schunk}
\begin{Soutput}
       expression    min median mem_alloc n_itr n_gc total_time
1      dplyr_mean  1.14s  1.25s   48.57MB     3   69      3.69s
2 data.table_mean  7.5ms 8.69ms   18.93MB   314   32      3.01s
3   collapse_mean 3.27ms 3.93ms    9.11MB   643   27         3s
\end{Soutput}
\begin{Soutput}
         expression     min  median mem_alloc n_itr n_gc total_time
1      dplyr_median   4.21s   4.21s    52.8MB     1   84      4.21s
2 data.table_median 21.92ms 23.71ms    18.9MB   122   13         3s
3   collapse_median  9.37ms 10.57ms    11.1MB   255   13      3.01s
\end{Soutput}
\end{Schunk}
So far, \pkg{data.table}, by virtue of it's internal vectorizations (also via dedicated grouped \proglang{C} implementations of simple functions), is competitive.\footnote{Much longer data will likely also favor \pkg{data.table} over \pkg{collapse} due to its sub-column-level parallel grouping and implementation of simple functions like \fct{sum} and \fct{mean}, see, e.g., the \href{https://duckdblabs.github.io/db-benchmark/}{DuckDB Benchmarks}.} Below, I compute the range of one column (\code{x}) using \code{max(x) - min(x)}. As elucidated in Section~\ref{sec:integration}, this expression is also vectorized in \pkg{collapse}, where it amounts to \code{fmax(x, g) - fmin(x, g)}, but not in \pkg{data.table}.
\begin{Schunk}
\begin{Sinput}
R> bmark(dplyr = flights |> group_by(month, day, origin, dest) |>
+      summarise(rng = max(arr_delay) - min(arr_delay), .groups = "drop"),
+    data.table = qDT(flights)[, .(rng = max(arr_delay) - min(arr_delay)),
+                              by = .(month, day, origin, dest)],
+    collapse = flights |> fgroup_by(month, day, origin, dest) |>
+      fsummarise(rng = fmax(arr_delay) - fmin(arr_delay)))
\end{Sinput}
\begin{Soutput}
  expression     min  median mem_alloc n_itr n_gc total_time
1      dplyr 91.78ms 107.7ms   20.18MB    27   55      3.14s
2 data.table 55.02ms  67.2ms    5.77MB    45   27      3.05s
3   collapse  5.17ms   5.7ms     6.8MB   491   14      3.01s
\end{Soutput}
\end{Schunk}
I also benchmark table joins and pivots. The following demonstrates how all tables can be joined together using \pkg{collapse} and its default first-match left-join, which preserves \code{flights}.
\begin{Code}
R> flights |> join(weather, on = c("origin", "time_hour")) |>
+    join(planes, on = "tailnum") |> join(airports, on = c(dest = "faa")) |>
+    join(airlines, on = "carrier") |> dim()
\end{Code}
\begin{Schunk}
\begin{Soutput}
left join: flights[origin, time_hour] 434526/435352 (99.8%) <21.94:1st> weat
duplicate columns: year, month, day, hour => renamed using suffix '_weather'
left join: x[tailnum] 424068/435352 (97.4%) <87.62:1st> planes[tailnum] 4840
duplicate columns: year => renamed using suffix '_planes' for y
left join: x[dest] 435352/435352 (100%) <3689.42:1st> airports[faa] 118/1255
left join: x[carrier] 435352/435352 (100%) <31096.57:1st> airlines[carrier]
duplicate columns: name => renamed using suffix '_airlines' for y
[1] 435352     48
\end{Soutput}
\end{Schunk}
The verbosity of \fct{join} is essential to understanding what has happened here---how many records from each table were matched and which duplicate non-id columns were suffixed with the (default) \code{y}-table name. Usually, I would set \code{drop.dup.cols = "y"} as keeping them is not helpful in this case, but the other packages don't have this option. For the benchmark, I set \code{verbose = 0} in \pkg{collapse} and employ the fastest syntax for \pkg{dplyr} and \pkg{data.table}:\footnote{\code{left\_join(..., multiple = "first")} for \pkg{dplyr} and \code{y[x, on = ids, mult = "first"]} for \pkg{data.table}.}
\begin{Schunk}
\begin{Soutput}
        expression      min median mem_alloc n_itr n_gc total_time
1      dplyr_joins 213.13ms  265ms   559.4MB    12   58      3.14s
2 data.table_joins 173.31ms  229ms     491MB    14   62      3.14s
3   collapse_joins   9.94ms   14ms    89.7MB   182  100      3.05s
\end{Soutput}
\end{Schunk}
Evidently, the vectorized hash join provided by \pkg{collapse} is 10x faster than \pkg{data.table} on this database, at a substantially lower memory footprint. It remains competitive on \href{https://duckdblabs.github.io/db-benchmark/}{big data}.\footnote{\pkg{data.table} joins utilize multithreaded radix-ordering---a very different logic more useful for big data.\vspace{-5mm}} \newline

Last but not least, I benchmark pivots, starting with a long pivot that simply melts the 6 columns aggregated beforehand into one column, duplicating all other columns 6 times:
\begin{Schunk}
\begin{Sinput}
R> bmark(tidyr = tidyr::pivot_longer(flights, cols = vars),
+        data.table = qDT(flights) |> melt(measure = vars),
+        collapse = pivot(flights, values = vars))
\end{Sinput}
\begin{Soutput}
  expression    min median mem_alloc n_itr n_gc total_time
1      tidyr 72.4ms  134ms     254MB    25   54      3.12s
2 data.table   43ms   59ms     209MB    45   34      3.11s
3   collapse 14.1ms   25ms     209MB    77   91      3.02s
\end{Soutput}
\end{Schunk}
Memory-wise, \pkg{collapse} and \pkg{data.table} are equally efficient, but \pkg{collapse} is faster, presumably due to more extensive use of \code{memset()} to copy values in \proglang{C} or smaller \proglang{R}-level overheads.

To complete the picture, I also also perform a wide pivot where the 6 columns are summed (for efficiency) across the 3 origin airports and expanded to create 18 airport-value columns.
\begin{Schunk}
\begin{Sinput}
R> bmark(tidyr = tidyr::pivot_wider(flights, id_cols = .c(month, day, dest),
+            names_from = "origin", values_from = vars, values_fn = sum),
+        data.table = dcast(qDT(flights), month + day + dest ~ origin,
+                           value.var = vars, fun = sum),
+        collapse_fsum = pivot(flights, .c(month, day, dest), vars,
+                              "origin", how = "wider", FUN = fsum),
+        collapse_itnl = pivot(flights, .c(month, day, dest), vars,
+                              "origin", how = "wider", FUN = "sum"))
\end{Sinput}
\begin{Soutput}
     expression      min   median mem_alloc n_itr n_gc total_time
1         tidyr 365.57ms 404.26ms     143MB     8   62      3.25s
2    data.table 222.76ms 231.33ms    21.7MB    13   44      3.08s
3 collapse_fsum   6.31ms   8.45ms    39.1MB   269   68         3s
4 collapse_itnl   4.03ms   5.19ms    12.4MB   513   38         3s
\end{Soutput}
\end{Schunk}
Again, \pkg{collapse} is fastest, as it offers full vectorization, either via \fct{fsum}, which translates to \code{fsum(x, g, TRA = "fill")} before pivoting and thus entails a full deep copy of the \code{vars} columns, or via the optimized internal sum function which sums values 'on the fly' during the reshaping process. \pkg{data.table} is not vectorized here but at least memory efficient. %\newline

In summary, these benchmarks show that \pkg{collapse} provides outstanding performance and memory efficiency on a typical medium-sized real-world database popular in the \proglang{R} community.
% %
% <<bench, cache=TRUE>>=
% setDTthreads(4)
% set_collapse(na.rm = FALSE, sort = FALSE, nthreads = 4)
% set.seed(101)
% m <- matrix(rnorm(1e7), ncol = 1000)
% data <- qDT(replicate(100, rnorm(1e5), simplify = FALSE))
% g <- sample.int(1e4, 1e5, TRUE)
%
% microbenchmark(R = colMeans(m), collapse = fmean(m))
% microbenchmark(R = rowsum(data, g, reorder = FALSE),
%                data.table = data[, lapply(.SD, sum), by = g],
%                collapse = fsum(data, g))
% add_vars(data) <- g
% microbenchmark(data.table = data[, lapply(.SD, median), by = g],
%                collapse = data |> fgroup_by(g) |> fmedian())
% d <- data.table(g = unique(g), x = 1, y = 2, z = 3)
% microbenchmark(data.table = d[data, on = "g"],
%                collapse = join(data, d, on = "g", verbose = 0))
% microbenchmark(data.table = melt(data, "g"),
%                collapse = pivot(data, "g"))
% settransform(data, id = rowid(g))
% cols = grep("^V", names(data), value = TRUE)
% microbenchmark(data.table = dcast(data, g ~ id, value.var = cols),
%           collapse = pivot(data, ids = "g", names = "id", how = "w"))
% @
%
%
% Apart from the raw algorithmic efficiency demonstrated here, \pkg{collapse} is often more efficient than other solutions by simply doing less. For example, if grouping columns are factor variables, \pkg{collapse}'s algorithms in \code{funique()}, \code{group()} or \code{fmatch()}, etc., use the values as hashes without checking for collisions. Similarly, if data is already sorted/unique, it is directly returned by functions like \code{roworder()}/\code{funique()}. \newline
%
\subsection{Other benchmarks}
The \href{https://duckdblabs.github.io/db-benchmark/}{DuckDB Benchmarks} compare many software packages for database-like operations using large datasets (big data) on a linux server. The January 2025 run distinguishes 6 packages that consistently achieve outstanding performances: \pkg{DuckDB}, \pkg{Polars}, \pkg{ClickHouse}, Apache \pkg{Datafusion}, \pkg{data.table}, and \pkg{collapse}. Of these, \pkg{DuckDB}, \pkg{ClickHouse}, and \pkg{Datafusion} are vectorized database (SQL) engines, and \pkg{Polars} is a Python/Rust DataFrame library and SQL engine. These four are supported by (semi-)commercial entities, leaving \pkg{data.table} as the only fully community-led project, and \pkg{collapse} as the only project that is single-authored and without financial support. The benchmarks show that \pkg{collapse} achieves the highest relative performance on 'smaller' (10-100 million row) datasets and performing advanced operations. \newline

Since June 2024, there is also an independent \href{https://github.com/AdrianAntico/Benchmarks?tab=readme-ov-file#background}{database-like operations benchmark by Adrian Antico} using a windows server and executing scripts inside IDEs (VScode, Rstudio), on which \pkg{collapse} achieved the overall fastest runtimes. I also very recently started a \href{https://github.com/fastverse/fastverse/wiki/Benchmarks}{user-contributed benchmark Wiki} as part of the \href{https://fastverse.github.io/fastverse/}{fastverse project} promoting high-performance software for R, where users can freely contribute benchmarks involving, but not limited to, \pkg{fastverse} packages. These benchmarks agree that \pkg{collapse} offers a computationally outstanding experience, particularly for medium-sized datasets, complex tasks, and on users PCs/Macs---which typically have smaller memory and parallel computing resources but faster chips than servers. % \footnote{Reasons for the particularly strong performance of \pkg{collapse} on Windows may be that it is largely written in C and has limited multithreading in favor or highly efficient serial algorithms---there appear to be persistent obstacles to efficient (low-overhead) multithreading on Windows, implying that multithreaded query engines do not develop their full thrust on medium-sized ($\leq$100 million row) datasets.}

\subsection{Limitations and outlook}

\pkg{collapse} maximizes three principal objectives: being class-agnostic/fully compatible with the \proglang{R} ecosystem (supporting statistical operations on vector, matrix and data.frame-like objects), being statistically advanced, and being fast. This warranted some design choices away from maximum performance for large data manipulation.\footnote{Which nowadays would demand creating a multithreaded, vectorized query engine with optimized memory buffers/vector types to take full advantage of SIMD processing as in \pkg{DuckDB} or \pkg{Polars}. Such an architecture is very difficult to square with \proglang{R} vectors and \proglang{R}'s 30-year old \proglang{C} API.} Its limited use of multithreading and SIMD instructions, partly by design constraints and by \proglang{R}'s \proglang{C} API, and the use of standard types for internal indexing, imposes hard-limits---the maximum integer in \proglang{R} is 2,147,483,647 $\to$ the maximum vector length \pkg{collapse} supports. It is and will remain an in-memory tool. \newline

Despite these constraints, \pkg{collapse} provides very respectable performance even on very large datasets by virtue of its algorithmic and memory efficiency. It is, together with the popular \pkg{data.table} package offering more sub-column-level parallel architecture for basic operations, well-positioned to remain a premier tool for in-memory statistics and data manipulation. % \newline

%% -- Summary/conclusions/discussion -------------------------------------------

\section{Conclusion} \label{sec:conclusion}

\pkg{collapse} was first released to CRAN in March 2020, and has grown and matured considerably over the course of 5 years. It has become a new foundation package for statistical computing and data transformation in \proglang{R}---one that is statistically advanced, class-agnostic, flexible, fast, lightweight, stable, and able to manipulate complex scientific data with ease. As such, it opens up new possibilities for statistics, research, production, and package development in \proglang{R}. \newline % It also sets an ambitious benchmark for high-quality software engineering on the language. \newline

This article provides a quick guide to the package, articulating its key ideas and design principles and demonstrating all core features. At this point the API is stable---it has changed very little over the 5 years and no further changes are planned. Compatibility with \proglang{R} version 3.5.0 will be maintained for as long as possible. Minor new features are currently planned. \newline

For deeper engagement with \pkg{collapse}, visit its \href{https://sebkrantz.github.io/collapse/index.html}{website} or start with the vignette summarizing all available \href{https://sebkrantz.github.io/collapse/articles/collapse_documentation.html}{documentation and resources}. Users can also follow \pkg{collapse} on \href{https://x.com/collapse_R}{Twitter/X} and \href{https://bsky.app/profile/rcollapse.bsky.social}{Bluesky} to be notified about major updates and participate in community discussions. \newline

Finally, \pkg{collapse} users are also encouraged to familiarize themselves with the \href{https://fastverse.github.io/fastverse/}{\pkg{fastverse}},\footnote{Website: https://fastverse.github.io/fastverse/} a suite of complementary high-performance packages for statistical computing and data manipulation in \proglang{R} that offer more advanced tools in several statistical computing domains. The \pkg{fastverse} metapackage additionally provides a lightweight framework to jointly load and manage these packages, as well as to build customized and fully separate package verses.

\newpage
%% -- Optional special unnumbered sections -------------------------------------

\section*{Computational details}
The results in this paper were obtained using \proglang{R} \citep{R} 4.4.3 with \pkg{collapse} 2.1.2, \pkg{data.table} 1.17.0, \pkg{dplyr} 1.1.4, \pkg{tidyr} 1.3.1, \pkg{matrixStats} 1.5.0, \pkg{fastverse} 0.3.4, \pkg{nycflights23} 0.2.0 \citep{rnycflights23}, and \pkg{bench} 1.1.4 \citep{rbench}. All packages used are available from the Comprehensive \proglang{R} Archive Network (CRAN) at https://CRAN.R-project.org/. The benchmark was run on an Apple MacBook Pro (2024) with a 48GB M4 Pro processor (single core speed $\sim$4.4 GHz). Packages were compiled from source using Apple Clang version 17.0.0 with OpenMP enabled and the -O2 flag. \newline

The \fct{bmark} function used for benchmarking is defined as follows:
\begin{Code}
bmark <- function(...) {
  bench::mark(..., min_time = 3, check = FALSE) |>
    janitor::clean_names() |>
    fselect(expression, min, median, mem_alloc, n_itr, n_gc, total_time) |>
    fmutate(expression = names(expression)) |>
    dapply(as.character) |> qDF()
}
\end{Code}

\section*{Acknowledgments}

The source code of \pkg{collapse} has been heavily inspired by (and partly copied from) \pkg{data.table} (Matt Dowle and Arun Srinivasan), \proglang{R}'s source code (R Core Team and contributors worldwide), the \pkg{kit} package (Morgan Jacob), and \pkg{Rcpp} (Dirk Eddelbuettel). Packages \pkg{plm} (Yves Croissant, Giovanni Millo, and Kevin Tappe) and \pkg{fixest} (Laurent Berge) have also provided a lot of inspiration (and a port to its demeaning algorithm in the case of \pkg{fixest}). I also thank many people from diverse fields for helpful answers on Stackoverflow and many other people for encouragement, feature requests, and helpful issues and suggestions.
% \begin{leftbar}
% All acknowledgments (note the AE spelling) should be collected in this
% unnumbered section before the references. It may contain the usual information
% about funding and feedback from colleagues/reviewers/etc. Furthermore,
% information such as relative contributions of the authors may be added here
% (if any).
% \end{leftbar}

%% -- Bibliography -------------------------------------------------------------
%% - References need to be provided in a .bib BibTeX database.
%% - All references should be made with \cite, \citet, \citep, \citealp etc.
%%   (and never hard-coded). See the FAQ for details.
%% - JSS-specific markup (\proglang, \pkg, \code) should be used in the .bib.
%% - Titles in the .bib should be in title case.
%% - DOIs should be included where available.

\newpage

\bibliography{refs}

%% -- Appendix (if any) --------------------------------------------------------
%% - After the bibliography with page break.
%% - With proper section titles and _not_ just "Appendix".

\newpage

\end{document}